\documentclass[journal]{IEEEtran}
\ifCLASSINFOpdf
\usepackage[pdftex]{graphicx}
\usepackage{float}
\usepackage{amsmath}
\usepackage{cite}
\usepackage{hhline}
\usepackage{multirow}
\usepackage{ctable}
\usepackage{array}
\newcolumntype{?}{!{\vrule width 1pt}}
\newcolumntype{x}[1]{>{\noindent\centering\hspace{0pt}}p{#1}}
\else
\fi

\begin{document}
	\title{Virtual Point Source Synthesis method for 3D Scintillation Detector Characterization}
	\author{Xin Li, Li Tao, Craig S. Levin and Lars R. Furenlid% <-this 
		\thanks{Manuscript received October 12, 2019.  This work was supported in part by National Institute of Health (NIH) under grants NIH/NIBIB 5-P41-EB002035-20 ``The Center for Gamma-Ray Imaging" and NIBIB 5R01EB000803-22 ``SPECT Imaging and Parallel Computing".\par}
		\thanks{X. Li and L. R. Furenlid are with the Center for Gamma-Ray Imaging, the Department of Medical Imaging and the College of Optical Sciences, University of Arizona, Tucson,
			AZ, 85721 USA. (email: xinli1@email.arizona.edu and furen@radiology.arizona.edu)}
		\thanks{L. Tao and C. S. Levin are with the Molecular Imaging Instrumentation Laboratory, Stanford University, Stanford, CA 94305 USA. (email: ltao2@stanford.edu and cslevin@stanford.edu)}
	}
	
	\markboth{ }%
	{Shell \MakeLowercase{\textit{et al.}}: Bare Demo of IEEEtran.cls for Journals}
	
	\maketitle
	
	\begin{abstract}
		A novel data-processing method was developed to facilitate scintillation detector characterization. Combined with fan-beam calibration, this method can be used to quickly and conveniently calibrate gamma-ray detectors for SPECT, PET, homeland security or astronomy. Compared with traditional calibration methods, this new technique can accurately calibrate a photon-counting detector, including DOI information, with greatly reduced time. The enabling part of this technique is fan-beam scanning combined with a data-processing strategy called the common-data subset (CDS) method, which was used to synthesize the detector's mean detector response functions (MDRFs). Using this approach, $2N$ scans ($N$ in x and $N$ in y direction) are necessary to finish calibration of a 2D detector as opposed to $N^2$ scans with a pencil beam. For a 3D detector calibration, only $3N$ scans are necessary to achieve the 3D detector MDRFs that include DOI information. Moreover, this calibration technique can be used for detectors with complicated or irregular MDRFs. We present both Monte-Carlo simulations and experimental results that support the feasibility of this method.
	\end{abstract}
	
	\begin{IEEEkeywords}
		Photon-counting, detector, calibration, depth of interaction (DOI), digital signal processing, SPECT, PET.
	\end{IEEEkeywords}
	
	\IEEEpeerreviewmaketitle

	\section{Introduction}
	Since its invention in 1952, the Anger camera and its derivatives have been widely used in gamma-ray imaging applications such as SPECT and PET, with ``Anger arithmetic" used as the standard method for gamma-ray interaction position estimation\cite{anger}\cite{angerBeyond}. However, this method exhibits large bias or variance in estimated positions, especially at areas close to detector edges. Reference-data based methods have been developed, which have better positioning performance over Anger arithmetic. These methods include, but are not limited to, maximum likelihood estimation (MLE)\cite{MLPR}, artificial neural networks (ANNs)\cite{ANN,rodNeuralNetWork,positionXenon,3DPositionNeuralNetWork,electronicsNNW,tao2020deep}, K-nearest neighbors (KNNs)\cite{KN} and k-d tree search\cite{kdt}. All of these methods require detectors to be calibrated to acquire reference data, which is subsequently used in gamma-ray interaction position estimation. The calibration or collection of reference data is usually performed using a well-collimated, normally-incident pencil beam of gamma rays to scan $N\times N$ positions across the entrance surface of the detector and collecting a fixed number of events at each of these positions\cite{fastSpectII}. This process can be very time-consuming, especially when the collimated beam diameter and step size are relatively small when calibrating high-resolution or large-area detectors. Newly proposed calibration methods include using two perpendicular fan beams\cite{borghi}; calibrating the detector with uncollimated gamma-ray sources after which algorithms are used to estimate the reference data\cite{floodC}; and generation of reference data by careful detector modeling\cite{detectorModeling}. Despite various available methods, the DOI calibration of thick photon-counting detectors still remains a challenge\cite{lerche2003depth,lerche2005depth,DOIDecoding}. A method that combines both efficiency and DOI capability is still needed. The introduced method, which has been briefly introduced in the preceding simulation work\cite{fastMonoSimulation}, takes off difficulty from the calibration process and places more burden on the data processing side, and it is capable of carrying out a 2D and 3D calibration (with DOI capability) of photon-counting detectors using two and three fan-beam scans, respectively. Moreover, by combining this DOI-enabled fast calibration methods with fast gamma-ray event position estimation methods such as k-d tree search \cite{kdt} or contracting-grid search \cite{CG}, broader application of monolithic gamma-ray detector might be possible.\par
	
	In this work, we proposed a fan-beam data processing method called the common-data subset (CDS) method, which could enable the DOI calibration of a monolithic crystal detector with reduced time. Both simulation and experiment are shown to validate this method.\par
	
	\section{Materials and methods}
	\subsection{Basis of common data subset (CDS) calibration}
	The basic idea of this calibration method is to use algorithms to find the events located at the overlapping positions of any intersecting fan beams.\par 
	\subsubsection{Two-dimensional (2D) calibration}
	As an example, a detector is first scanned by a vertical fan beam moving along x direction; and then a horizontal beam is used to scan the detector along y direction. For both horizontal and vertical directions, the fan beam stops at $N$ positions and reference data is collected at each of the positions (Fig. 1).\par
	
	\begin{figure} [H]
		\begin{center}
			\includegraphics[scale=0.3]{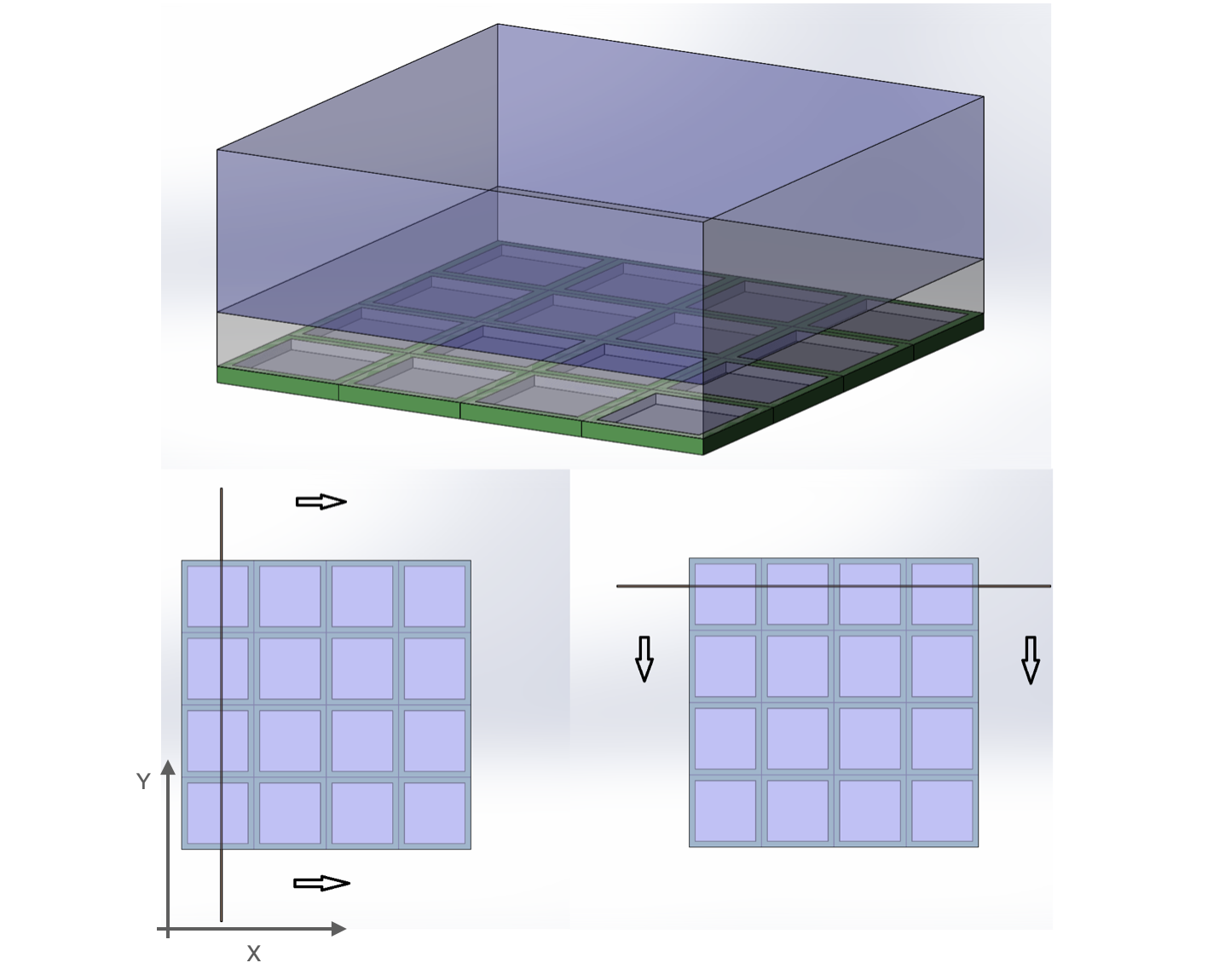}
			\caption{Fan beams scanning a detector (composed of a monolithic crystal, a light guide and $4\times4$ array of SiPMs) along two orthogonal directions. The scanning fan beam stops at $N$ positions along both horizontal and vertical directions, and reference dataset is collected at each of the positions.}
		\end{center}
	\end{figure}\par
	
	$N$ scans are collected in each of the two directions as shown in Fig. 1. Each of the $N$ scans includes certain number of gamma-ray interaction events, which form a reference dataset, therefore there are in total $2N$ reference datasets collected as Fig. 1 shows. In this work, we define that an event $\mathbf{a_i}$ in one dataset $\mathbf{A_n}$ represents the signals of all the detector's light sensors produced by a single gamma-ray interaction (1).\par
	
	\begin{equation}
		\begin{aligned}
			& \mathbf{a_i\in A_n}\\
			\mathbf{a_i = \{s_1}&\mathbf{, s_2, s_3,..., s_M\}}	
		\end{aligned}
	\end{equation}
	
	$\mathbf{i}$ is the index of a single event in a reference dataset, and $n=1, 2, 3,..., N$ is the index of datasets (for example, index of the $N$ datasets along x direction, Fig. 1). Assuming that the detector has a total of $M$ light sensors, then each reference event (vector) has $M$ signal components. Two gamma-ray interactions that occur in close proximity inside the detector will give two similar signal responses. We assert the opposite is also true, namely that:\\
	{\bf If two gamma-ray interactions yield similar signal responses, their interaction positions in the photon-counting detector are close to each other}.\\
	This assumption is true if the detector is well-designed so that the interaction positions and their mean signal responses have one-to-one mapping relationships, while a poorly-designed detector has the same signal responses even for events interacting at different positions, thus causing position-estimation ambiguity when using the signal response of an event to estimate the event's position of interaction. Suppose there are two calibration datasets $\mathbf{A_n}$ and $\mathbf{B_{n'}}$ acquired by two orthogonal fan beams. Then we search for the data subsets $\mathbf{C_{An} \in A_{n}}$ and $\mathbf{C_{Bn'} \in B_{n'}}$ such that $\mathbf{C_{An} \approx C_{Bn'}}$ (meaning that the two event subsets have similar events compared with each other). Dataset $\mathbf{C_{nn'} = C_{An} \cup C_{Bn'}}$ is defined to be the common-data subset (CDS) of datasets $\mathbf{A_n}$ and $\mathbf{B_{n'}}$. Considering the above assumption, the common-data subset $\mathbf{C_{nn'}}$ contains the gamma-ray events that interact inside or close to the overlapping volume of the two calibration fan beams.\par
	
	If the CDS of two datasets can be found using algorithms, the calibration datasets could effectively generate what would be produced by a pencil beam (Fig. 2). \par
	
	\begin{figure} [H]
		\begin{center}
			\includegraphics[scale=0.25]{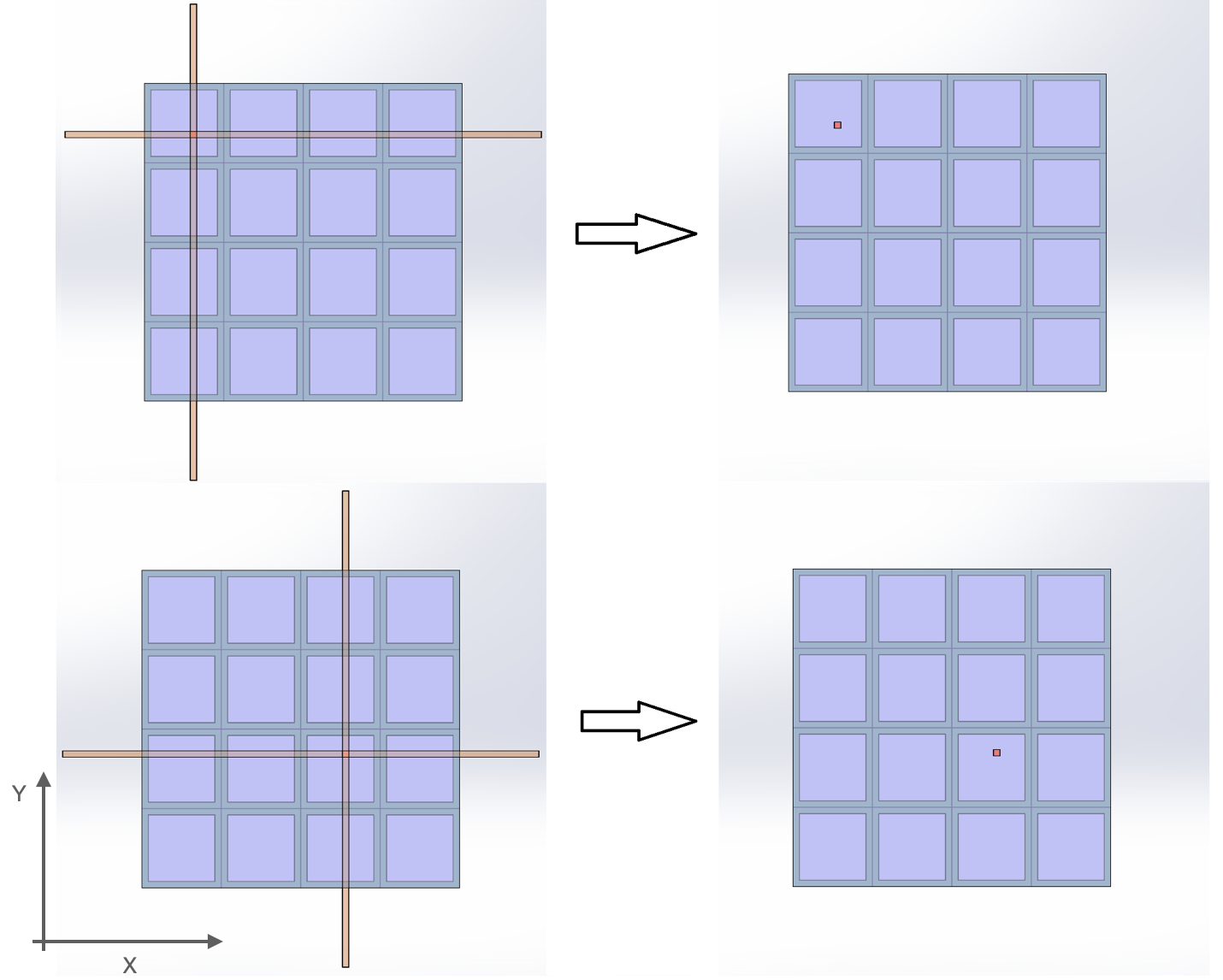}
			\caption{Common data subset (CDS) can be found using algorithms to process reference datasets collected with fan beams. The CDS is similar to the dataset produced by a pencil beam illuminating the overlapping region of the two fan beams.}
		\end{center}
	\end{figure}\par
	
	After all of the reference datasets of the fan beam scanning horizontally are acquired and represented as \{$\mathbf{A}$\} = \{$\mathbf{A_1, A_2,...,A_N}$\}, and all the reference datasets of the fan beam scanning vertically are also acquired and represented as \{$\mathbf{B}$\} = \{$\mathbf{B_1, B_2,...,B_N}$\}, different combinations of $\mathbf{A_n}\in \{\mathbf{A}\}$ and $\mathbf{B_{n'}} \in \{\mathbf{B}\}$ yield an $N\times N$ grid of calibration datasets that can be used to generate the detector signal responses or mean detector response functions (MDRFs) -- the mean signal map of each light sensor  of the detector, as a function of gamma-ray interaction position in the detector volume.\par
	
	Using this method, only $2N$ fan-beam scans are necessary, that greatly cut down the calibration time, as opposed to that of $N^2$ scans using a pencil beam. Making a well-collimated fan beam is much easier than making a well-collimated pencil beam, especially for high-energy gamma rays.\par
	
	\subsubsection{Three-dimensional (3D) detector calibration}
	
	Analogously to the 2D detector case, in order to calibrate the 3D signal responses (3D MDRFs) for a thick detector, only $3N$ scans by a fan beam are necessary, as illustrated in Fig. 3.
	
	\begin{figure} [H]
		\begin{center}
			\includegraphics[scale=0.175]{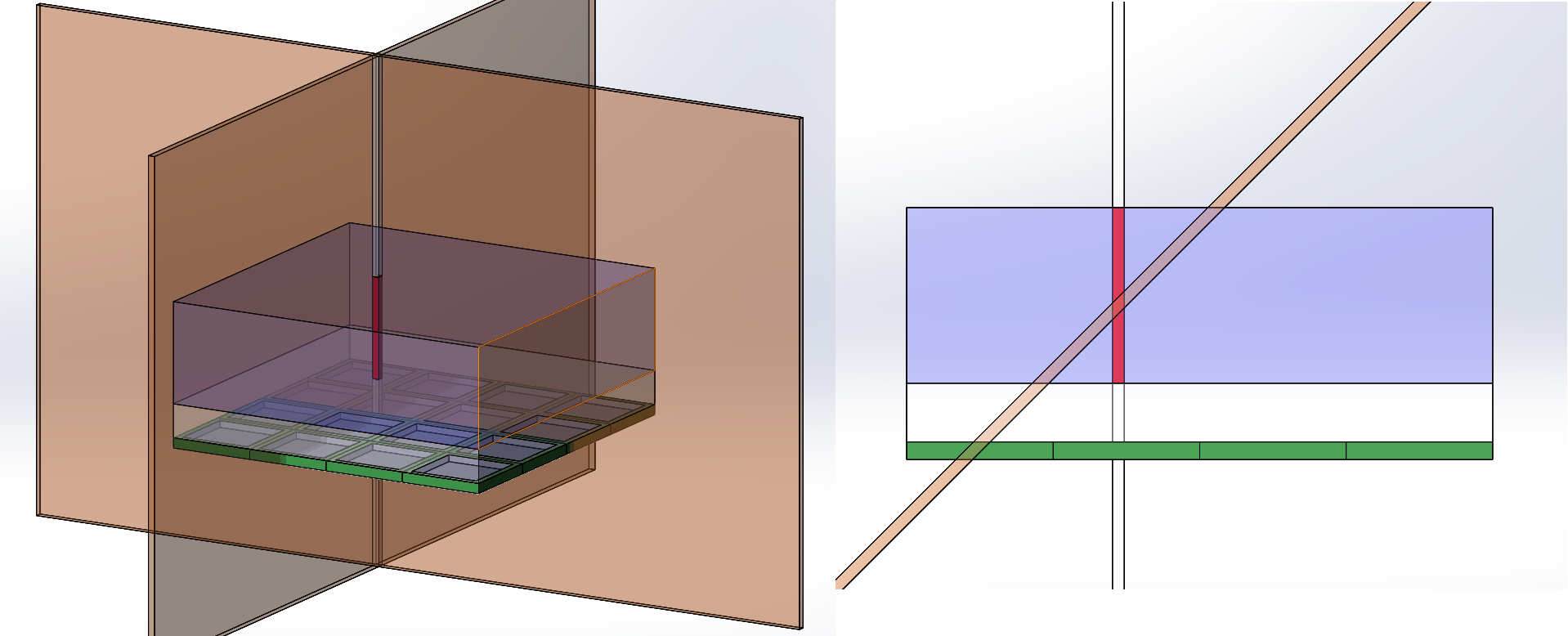}
			\caption{Three fan-beam calibration scans can be used to calibrate a 3D detector, such as a scintillation detector with very thick crystal. The first and second beams are similar to those of the 2D case, and the third scan could be done with a slant beam, with an incident angle (for example, $\approx45^o$) between the calibrating fan beam and the detector's entrance surface. The reason to use a slant beam, instead of using a beam parallel to the entrance surface, is to avoid extreme long travel distance of beam photons inside the detector volume, which will cause severe attenuation of the beam when detector area is large.}
		\end{center}
	\end{figure}\par
	
	The first two orthogonal fan beams are used to produce the CDSs whose corresponding interaction locations are located at or close to the overlapping detector volume of the two fan beams, which defines a tubular volume (red area in Fig. 3). The third calibrating beam is used to intersect the tubular segment to define a small volume (similarly, the intersection of three planes is a point). By translating the three fan beams, the whole 3D detector volumes can be calibrated, then the 3D MDRFs can be derived based on different combinations of the $3N$ achieved datasets.\par
	
	In this case, the DOI calibration can be performed easily and conveniently. DOI estimation could help PET system to improve radical spatial resolution close to edges of field of view  (FOV) and help pin-hole SPECT system to reduce parallax errors near edges of detectors. In most cases, the third fan beam can be orthogonal to the first and second fan beams which is often preferred due to its simplicity. However, for some specific scenarios, using a slant third beam is more beneficial. For example, if a monolithic SPECT detector has a large and flat scintillation crystal (large entrance surface, with small thickness), a third beam orthogonal to the first and second beams will be perpendicular to the edge surface of the scintillation crystal, and its travel distance inside the scintillation crystal will be relatively long, leaving very few interaction events where the beam exits the crystal, thus the statistic noise will be higher. If using a slant beam, the attenuation effect would be greatly mitigated due to a shorter travel distance inside the crystal.
	
	\subsection{Algorithm for finding the common data subset (CDS) of two datasets}
	
	The basic mathematical task is to efficiently find the common data subset (CDS) of two datasets under the influence of noise. Virtually no event (signal vector) in dataset $\mathbf{A_n}$ will be exactly identical to any event in dataset $\mathbf{B_{n'}}$. However, the events in dataset $\mathbf{A_n}$, that are similar to one or more events in dataset $\mathbf{B_{n'}}$ could be found out.
	
	\begin{equation}
		\begin{aligned}
			\mathbf{C_{A_n}}=\{\mathbf{a_i} \in \mathbf{A_n}\hspace{0.2cm}|\hspace{0.2cm} \exists \mathbf{b_j} \in \mathbf{B_{n'}} \hspace{0.2cm}so\hspace{0.2cm} \mathbf{b_j} \approx \mathbf{a_i} \}	
		\end{aligned}
	\end{equation}
	
	And similarly, $\mathbf{C_{B_{n'}}}$ will be found according to (3). 
	
	\begin{equation}
		\begin{aligned}
			\mathbf{C_{B_{n'}}}=\{\mathbf{b_j} \in \mathbf{B_{n'}}\hspace{0.2cm}|\hspace{0.2cm} \exists a_i \in \mathbf{A_n} \hspace{0.2cm}so\hspace{0.2cm} \mathbf{a_i} \approx \mathbf{b_j} \}	
		\end{aligned}
	\end{equation}
	
	Then the common data subset (CDS) can be formed by taking the union of the above two datasets.
	
	\begin{equation}
		\begin{aligned}
			\mathbf{C_{nn'}}=\mathbf{C_{A_n}}\cup \mathbf{C_{B_{n'}}}
		\end{aligned}
	\end{equation}
	
	To quantify if an event in dataset $\mathbf{A_n}$ is ``similar" or ``close" to a particular event in dataset $\mathbf{B_{n'}}$, a score should be defined. Because each event's signal vector has $M$ components (where $M$ represents the number of light sensors or readout channels of the detector), each event is a signal vector $\vec{\mathbf{s}}=\{\mathbf{s_1, s_2,...,s_M\}}$. The closeness of an event $\mathbf{a_i}$ in dataset $\mathbf{A_n}$ to dataset $\mathbf{B_{n'}}$ can be, for example, quantified by the following score equation. 
	
	\begin{equation}
		\begin{aligned}
			score(\mathbf{a_i}, \mathbf{B_{n'}})=\sum_{\mathbf{b_j}\in \mathbf{B_{n'}}} {e^{-\frac{\sum_{k=1}^{M}(\mathbf{a_{ik}} - \mathbf{b_{jk}})^2}{c}}}.
		\end{aligned}
	\end{equation}
	
	Where $c$ in the above equation is a tunable constant, which represents the radius of the field extended by each reference event in $\mathbf{B_{n'}}$. Each event in the reference dataset extends a field in $M$ dimensional space, and the summation of all the field values in a specific location (decided by the query event) in the $M$ dimensional space defines the score value of that event. For the above score definition, the shape of field is Gaussian. This $c$ value needs to be tuned, since if $c$ is too large, the resolution of the CDS method will be poor, so events outside the intersection volume of the two beams might be mixed into the common-data subset (CDS). But if $c$ is too small, the radius of the field extended by each reference event will be too small, so the score of an event located at the common volume will sometimes have very small score value, which could be excluded from the CDS; while an event outside the common volume could be wrongly included into the CDS if it happens to have a higher score value under the influence of noise. This example shows only one of many available options to define the score, other definitions could be possible, such as changing the power of 2 in equation (5) to 1, 3, or 4...\par 
	
	In order to find the CDS of two datasets $\mathbf{A_{n}}\in \mathbf{A}$ and $\mathbf{B_{n'}}\in \mathbf{B}$: first, calculate the score of $\mathbf{a_i}$ across all the datasets $\mathbf{B_{n'}}\in \mathbf{B}$; find the dataset in $\mathbf{B}$ that gives the highest score; then assign $\mathbf{a_i}$ to that dataset, and the event $\mathbf{a_i}$'s coordinate of the direction along which datasets $\mathbf{B}$ are collected can be achieved. Repeat the same procedure for other directions, then the 3D coordinate of event $\mathbf{a_i}$ can be achieved. All events in the reference datasets can be processed in this way, and we call this method the ``competitive method". By default, we will use this method for the rest of this work.\par
	
	Once the CDSs are achieved, the events in the reference datasets can be associated with their interaction positions in the detector volume, and the corresponding MDRF value, namely $\mathbf{MDRF_k(x,y,z)}$, can be calculated easily by averaging the signal vectors of all the events in the CDS of location $(x,y,z)$, where $k$ is the index of light sensors. \par 
	
	After calculating the MDRFs, the estimated location $\mathbf{\hat{x_i},\hat{y_i},\hat{z_i}}$ of a query event $\mathbf{q}$ can be found using maximum likelihood position estimation\cite{MLPR}.\par
	
	\begin{equation}
		\begin{aligned}
			\mathbf{\hat{x_i},\hat{y_i},\hat{z_i}}&=\underset{x, y, z}{argmax}\sum_{k=1}^{M} ln(prob(q_k|\mathbf{MDRF_k(x,y,z)})\\
			\mathbf{q}&= {\{q_1, q_2, q_3,..., q_M\}}	
		\end{aligned}
	\end{equation}
	
	\subsection{Simulation}
	
	\subsubsection{Simulation of a thick monolithic PET detector}
	The geometry of a simulated detector is shown in Fig. 4. The detector is composed of a thick LYSO scintillation crystal ($26$mm$\times 26$mm$\times 20$mm, refractive index is $1.85$), a light guide made of glass ($26$mm$\times 26$mm$\times 5$mm, refractive index is $1.55$), and $16$ SiPMs of dimensions $6.5$mm$\times 6.5$mm with sensitive area of $6$mm$\times 6$mm, forming a $4\times 4$ array on bottom of the light guide. The window material of the SiPMs has a thickness of $500$um and a refractive index of $1.52$. The photon detection efficiency (PDE) of the SiPMs was modeled to be $25$\%. The SiPMs' dark count rate is $1.0$MHz/mm$^2$, and excess noise factor (ENF) was modeled to be $1.21$ to avoid overly-optimistic result. All the parameters of the simulated SiPMs were chosen conservatively within the scope of current industrial capability \cite{hamamatsu, sensl}. The entrance surface of the detector is a Lambertian reflector, with reflectance of $95$\%, while the side walls of the detector are black absorber, with reflectance of $5$\%. Optical gel of refractive index of $1.49$ was modeled as the optical coupling medium between the light guide and the SiPMs, and its thickness is $20$um. The surfaces of the SiPMs were modeled to be specular surfaces with $40$\% reflectance. The LYSO crystal's light yield was modeled to be $38,000$ photons/MeV\cite{saintgobain}, with its intrinsic energy resolution modeled to be $9.5$\%, its attenuation length ($1/e$) for $511$ keV gamma rays is $12$mm, while its optical attenuation length ($1/e$) is $40$cm\cite{attLength}. The simulation is done using C++, and data processing of CDS is done using C++/CUDA.\par 
	
	In order to study the impact of Compton effect, we first simulated without considering Compton effect by letting each interacting gamma-ray photon interact by photoelectric effect. After that, we carried out another simulation including Compton effect. At last, we compared the results with and without Compton effect. The flow chart of how each gamma-ray photon is simulated is shown in Fig. 5. Each gamma ray is modeled as follows:\par
	1. Each gamma-ray photon can be described by its positon $x, y, z$, its direction $v_x, v_y, v_z$, and its energy $E$. The energy of a single gamma-ray photon is used to get its corresponding Compton scattering and photoelectric cross sections ($\sigma_{Compton}$ and $\sigma_{photo}$), and the total interaction cross section $\sigma_{total}$ is the sum of Compton scattering and photoelectric cross sections. \par
	2. A random number is used together with the total interaction cross section to determine how far this gamma-ray photon can travel before interacting with the scintillation material. Then the position of this gamma-ray photon $x, y, z$ is updated with this new interaction position.\par
	3. Another random number is used to determine whether this gamma-ray photon interacts with scintillator by Compton or photoelectric effect. If the gamma-ray photon interacts by Compton effect, the scattering angle of scattered photon is sampled from the pre-stored look-up table (LUT) calculated using Klein-Nishina formula \cite{klein2014scattering}. The LUT is a two-dimensional matrix, with its row axis indicating the incident gamma ray's energy, and the column axis indicating the scattering angle. The value in the matrix indicates the probability of each scattering angle, given a gamma ray's incident energy, so each row of the matrix represents the scattering angle's probability density function. After the scattering angle is achieved by sampling from the LUT, the azimuth angle is sampled uniformly from $0$ to $2\pi$, then the direction vector of the scattered gamma-ray ($v_x, v_y, v_z$) is fully determined, and the scattered gamma-ray photon's energy is calculated using the Compton formula. The energy deposited locally by Compton effect is then known, which is used to calculate the number of emitted visible photons by the scintillator, and each visible photon is traced until it is either detected or absorbed internally without being detected. Then go back to step 1, and repeat the whole process for the scattered gamma-ray photon.\par
	4. If the gamma-ray photon interacts by photoelectric effect, all of its energy is deposited locally at the interaction position, and visible photons are emitted. Each visible photon is traced until it is either detected or absorbed internally without being detected, then the simulation of this particular gamma-ray photon is done.\par
	\begin{figure} [H]
		\begin{center}
			\includegraphics[scale=0.30]{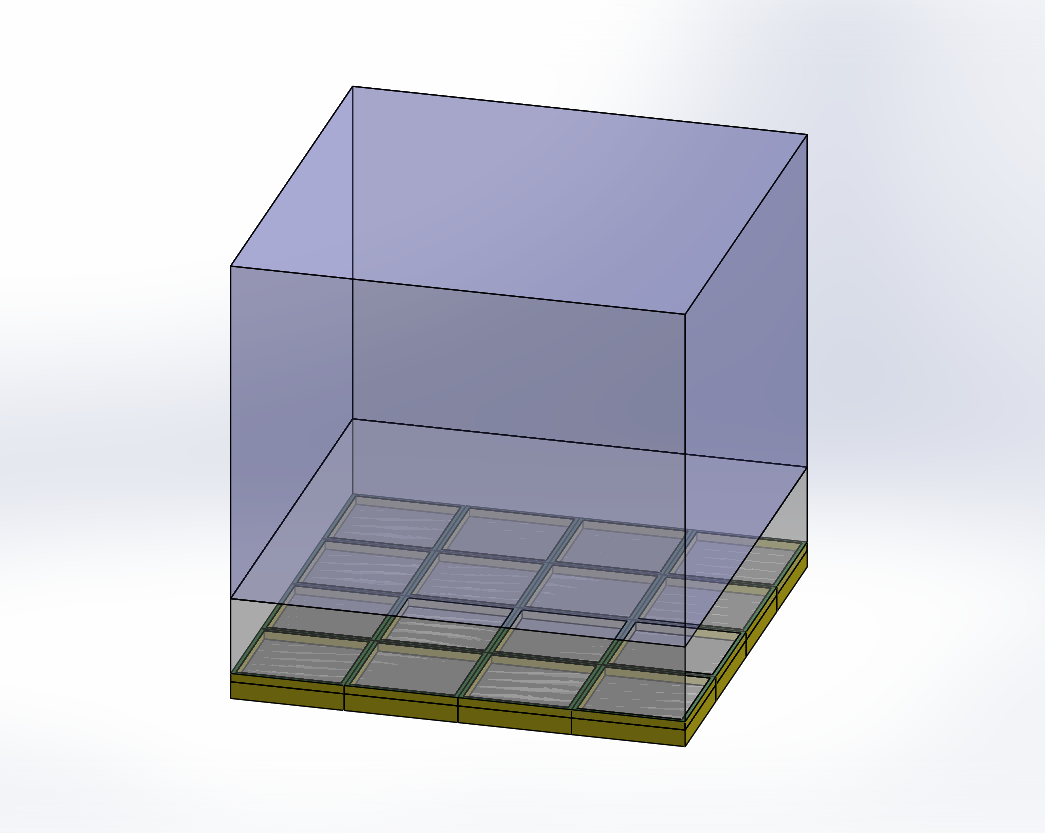}
			\caption{The rendered picture of the simulated thick monolithic crystal detector, the light guide is the layer of glass (grey) sandwiched between the scintillation crystal (blue) and SiPM array.}
		\end{center}
	\end{figure}\par
	
	\begin{figure} [H]
		\begin{center}
			\includegraphics[scale=0.21]{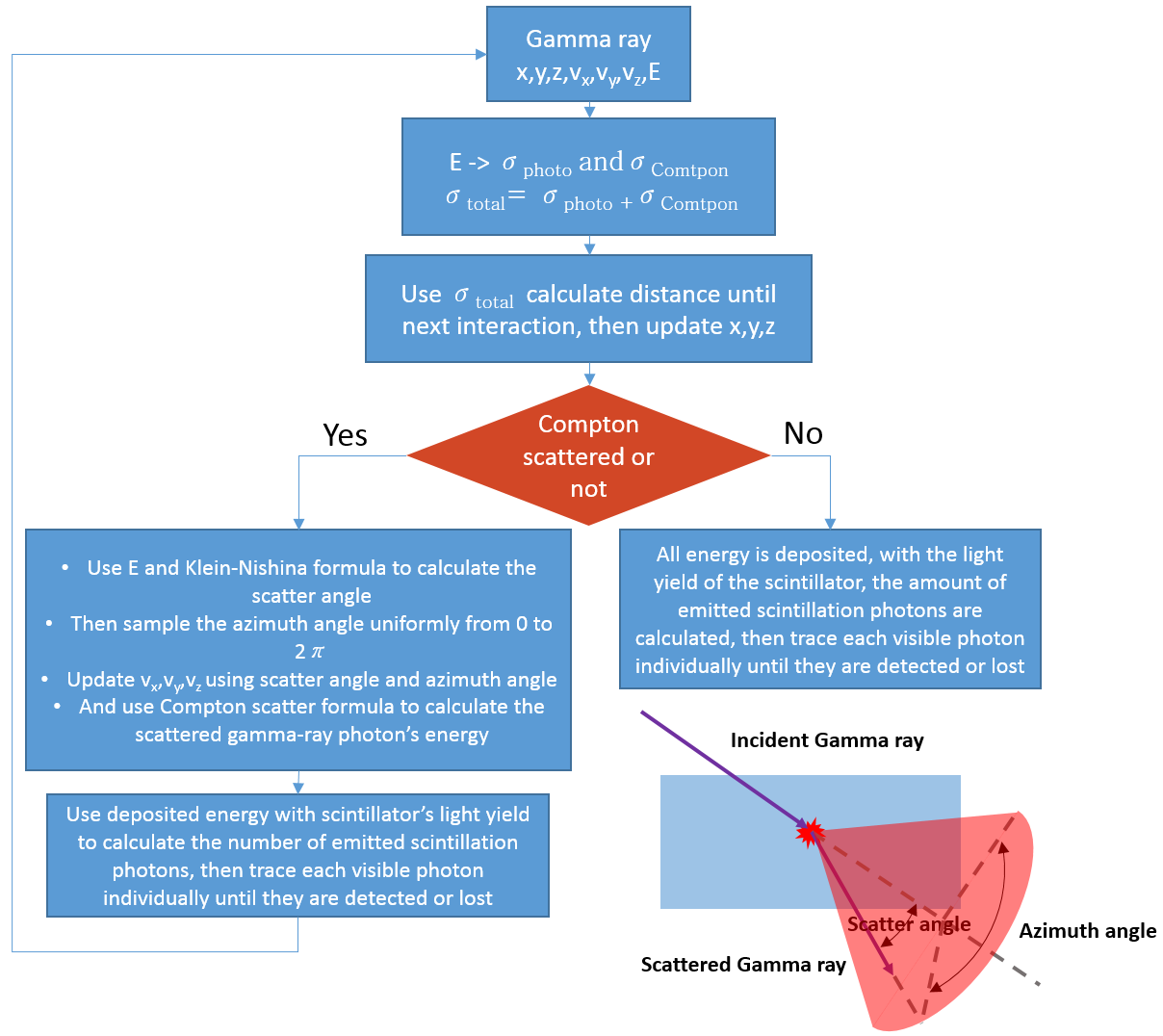}
			\caption{The flow chart showing the simulation process including Compton effect.}
		\end{center}
	\end{figure}\par
	
	Three fan beams are simulated to scan across the detector, as shown in Fig. 6.\par
	
	\begin{figure} [H]
		\begin{center}
			\includegraphics[scale=0.20]{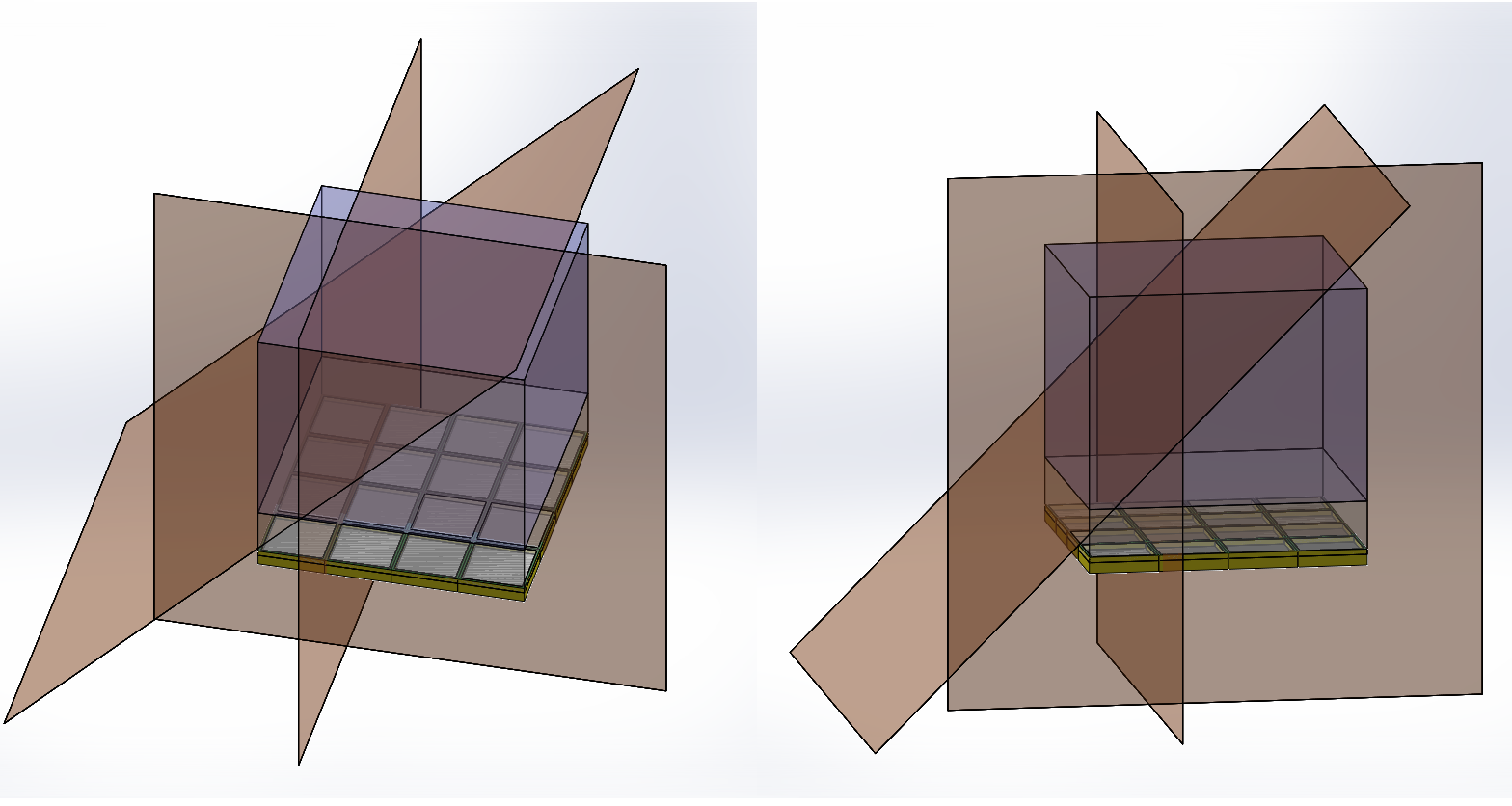}
			\caption{The scanning fan beams and the simulated monolithic crystal detector. Gamma-ray photons are collimated to form fan beams, which are used to scan the monolithic detector sequentially.}
		\end{center}
	\end{figure}\par
	
	A fan beam ($0.2355$mm FWHM) scanned the detector along x direction, then the fan beam was rotated $90\deg$, and scanned the detector along y direction. After that, the beam is inclined at $45^{\circ}$ angle with respect to entrance surface of the detector and scanned the detector along x direction. The fan beam each stopped at $52$ locations along the scintillation crystal when scanning along x and y direction, with a step size of $0.5$mm, starting/ending at $0.25$mm (half pitch size) from the side surfaces of the crystal. When scanning with a slant angle, the fan beam stopped at $92$ locations, with step size of $0.5$mm along x direction. At the starting point, the slant fan beam intersected the side surface of the scintillation crystal at $0.25$mm from the bottom surface. At the ending point, the slant beam intersected the crystal's top surface at $0.25$mm from the side surface. For simulation excluding Compton effect, at each scanned location, $50,000$ of $511$ keV gamma photons were emitted, while for simulation including Compton effect, at each scanned location, $150,000$ events were emitted to compensate for the loss after applying an energy window to filter out Compton-scattered events that deposit partial energy. The detected events at each scanned location form a dataset. So there were $52$ datasets formed by the fan beam scanning along x direction, $52$ datasets formed by the fan beam scanning along y direction, and $92$ datasets formed by the slant fan beam scanning along x direction. For each gamma-ray interaction inside the scintillation crystal, a burst of visible scintillation photons was emitted. The number of emitted visible photons was calculated by the amount of energy deposited, the light yield of scintillation crystal and the intrinsic energy resolution of scintillation crystal. Each visible photon was traced until it was either detected or absorbed without being detected.\par
	
	After the datasets were achieved, the MDRFs of the scintillation detector were calculated using the CDS method introduced in section II, part B. That is: for each event, find the dataset with highest score among all the datasets scanning the detector along a particular direction (x, y or  slant beam scanning x direction). Then the coordinate of this event can be decided. The same procedure was repeated for all the events, and the MDRF values associated with a particular voxel can be calculated by averaging all the signal vectors of events in that voxel.\par
	
	The above process was repeated for the simulation including Compton effect, to see if the CDS method fails under the influence of Compton effect.\par
	
	After the MDRFs were calculated with the CDS method, another set of MDRFs achieved from scanning a point source inside the detector was used for comparison (ground truth). $250$ photoelectric gamma-ray interactions were recorded at each of the $52\times 52\times 40$ grid points, and the same dataset was also used as test events to compare each event's true position and estimated position using MDRFs, to access the accuracy of the MDRFs.\par
	
	\subsubsection{Simulation of an edge-readout detector}
	
	This time, the simulated detector is the edge-readout detector\cite{edge} (Fig. 7). Its scintillation material was modeled to be CsI(Tl) of dimension $27.4$mm$\times27.4$mm$\times3$mm, the light yield of CsI(Tl) crystal was modeled as $54,000$ photons/MeV\cite{saintgobain}. $16$ SiPMs were assembled on the four edges of the CsI(Tl) crystal layer, and $4$ pieces were attached on each edge. The dimension of each SiPM was $7.35$mm$\times6.85$mm with active area of $6.0$mm$\times6.0$mm. Since the SiPM active area's height is $6.0$mm, and CsI(Tl) crystal's thickness is $3.0$mm, about half of each SiPM's active area was not used. The simulation geometry is the same as that in the experiment section (section II, part D).\par
	A $4\times4$ grid of drilled-hole optical barriers was simulated\cite{edge}. Each hole's surface was modeled as diffusive cylindrical surface which is semitransparent with absorption coefficient of $20$\%, its diameter is $1.08$mm and height is $3$mm . The distance between two adjacent optical barriers is $5.82$mm.\par
	The photon detection efficiency (PDE) of the SiPMs was modeled as $20$\% (considering the mismatch between the peak PDE wavelength of SiPM and CsI(Tl) peak emission wavelength), and the reflectance of the SiPMs was modeled as $40$\%. Between the scintillation crystal and the SiPMs, a $20$um-thick layer of optical gel with refractive index of $1.49$ was modeled. The dark count rate of each SiPM was assumed to be $1.0$ MHz/mm$^2$. The excess noise factor (ENF) was modeled as $1.21$. Again, all the SiPM parameters in the simulation were chosen conservatively according to the major manufacturers' data sheet \cite{hamamatsu, sensl}. The simulation was done using C++, and CDS data processing is done using C++/CUDA.\par
	Two orthogonal fan beams ($0.625$mm FWHM) were modeled to scan the 2D detector along x and y directions sequentially across an area of $20.0$mm$\times20.0$mm. $81$ locations were scanned for each direction, with a step size of $0.25$mm. The energy of the gamma-ray beam was modeled as $662$ keV (produced by Cs-137) and $50,000$ gamma-ray interactions were launched at each of the scanned locations. Again, the simulation parameters were chosen to be the same as the experiment section (section II, part D).\par
	
	\begin{figure} [H]
		\begin{center}
			\includegraphics[scale=0.28]{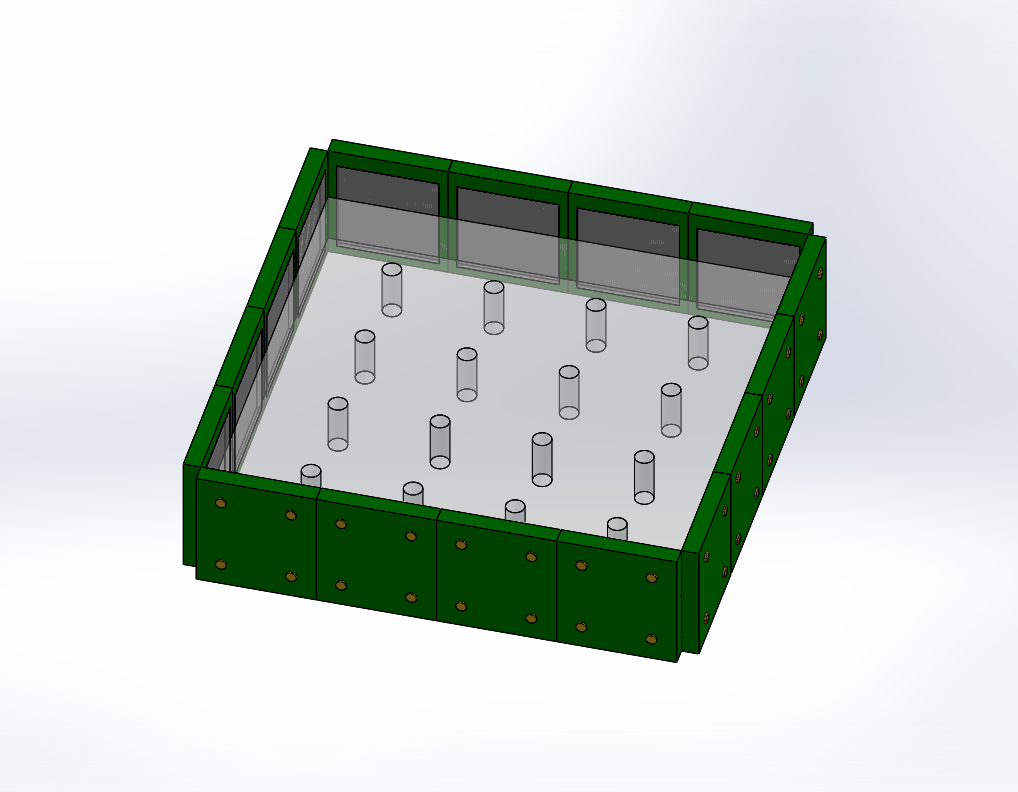}
			\caption{A rendering of the simulated single-layer edge-readout detector, with $16$ SiPMs attached to the edges of a CsI(Tl) crystal; $16$ drilled holes were added as optical barriers\cite{edge}.}
		\end{center}
	\end{figure}\par
	
	The 2D MDRFs of the edge-readout detector were calculated using the CDS method introduced in section II, part B. After that, another set of MDRFs were simulated by scanning a Gaussian beam of $662$ keV photons (FWHM is $0.625$mm) across $81\times 81$ locations inside the edge-readout detector, and this set of MDRFs were used for comparison. The set of events generated with this Gaussian beam was also used as test events, whose real interaction positions were later compared with estimated interaction positions achieved using MDRFs.\par
	
	\subsection{Experimental verification}
	
	The experiment used the edge-readout detector with drilled-hole optical barriers\cite{edge}, whose geometry is the same as the above simulation section (section II, part C), with the width of the fan beam measured to be $0.625$mm FWHM. The calibration was performed by scanning the detector at $81$ positions along horizontal direction. After that, the beam was rotated by $90^{\circ}$ and another $81$ scans were collected along vertical direction. The step size was $0.25$mm, with $50,000$ events with full energy deposition collected at each position. The energy window was chosen to be $\pm 12$\%\cite{edge}.  The experiment setup is shown in Fig. 8.\par
	
	\begin{figure} [H]
		\begin{center}
			\includegraphics[scale=0.3]{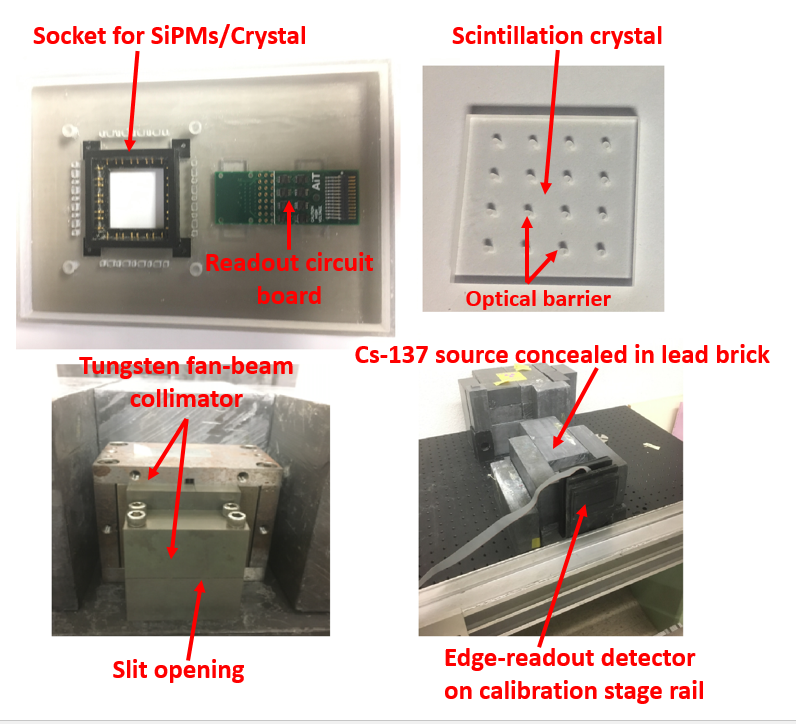}
			\caption{Experimental setup of the edge readout detector. Top-left: the detector socket with preamplifier circuit. Top-right: CsI(Tl) crystal with drilled-hole optical barriers. Bottom-left: tungsten slit collimators used to make the fan beam and crossed-slit pencil beam. Bottom-right: the assembled edge-readout detector (optically concealed by black tape) being scanned on a 2D stage system.}
		\end{center}
	\end{figure}\par
	
	%\begin{figure}[!t]
	%\centering
	%\includegraphics[width=2.5in]{myfigure}
	% where an .eps filename suffix will be assumed under latex, 
	% and a .pdf suffix will be assumed for pdflatex; or what has been declared
	% via \DeclareGraphicsExtensions.
	%\caption{Simulation Results.}
	%\label{fig_sim}
	%\end{figure}
	
	After collection of the datasets, the CDS method (section II, part B) was used to generate the 2D MDRFs. The CDS-generated MDRFs were then compared with the MDRFs measured using the conventional pencil beam method, which was realized by crossed-slits method using the same slits making the fan beams, which gave a spot size of $0.625$mm$\times0.625$mm FWHM at detector surface.\par
	
	The event data achieved by a $662$ keV pencil beam created with crossed slits has a high variance due to aperture leakage and scattering. MDRFs are therefore calculated via an iterative filtering process at each scanning position. The steps are: 1) the mean of the signal vector is calculated; 2) each event vector's Euclidean distance to the mean is calculated; 3) the events with the largest $5$\% Euclidean distance are eliminated; 4) loop back to step 1. This process is repeated until the result converges, for example, each events' Euclidean distance to mean is lower than a preset threshold. At convergence, the fraction of the events contained in the calculation of the MDRF value at each scanning position is around $20$\% of the initial number of events. However, the iterative filtering method could introduce bias at the place where the MDRF gradient is high (like positions close to optical barriers).\par
	
	In order to make a thorough comparison of the CDS and pencil-beam calibration methods, the events of several slit beams at different positions were estimated using the MDRFs achieved by the two methods. $5$ horizontal and $5$ vertical line images were estimated using maximum-likelihood method, each of the $5$ horizontal and $5$ vertical lines has a beam width of $0.625$mm and contains $50,000$ gamma-ray interaction events (full energy deposition). The lines were equally-spaced with a spacing of $4.0$mm. The central line is aligned with the center of the detector.
	
	\section{Results}
	\subsection{Simulation of a thick monolithic detector}
	
	The 3D MDRFs generated by scanning a simulated point light source inside the scintillation crystal is shown in Fig. 9. Only three sets of MDRFs at three different depths are shown. And these MDRFs are used as ground truth.
	
	\begin{figure} [H]
		\begin{center}
			\includegraphics[scale=0.32]{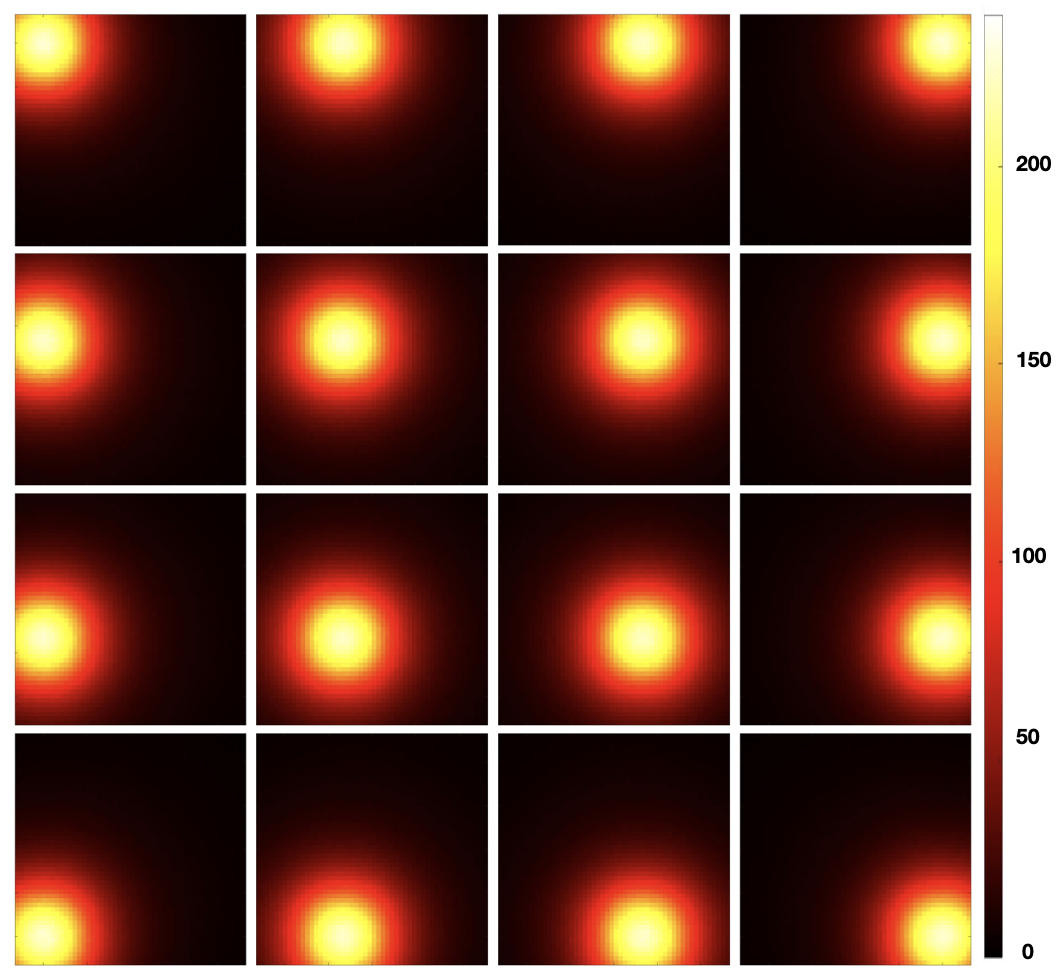}
			\includegraphics[scale=0.32]{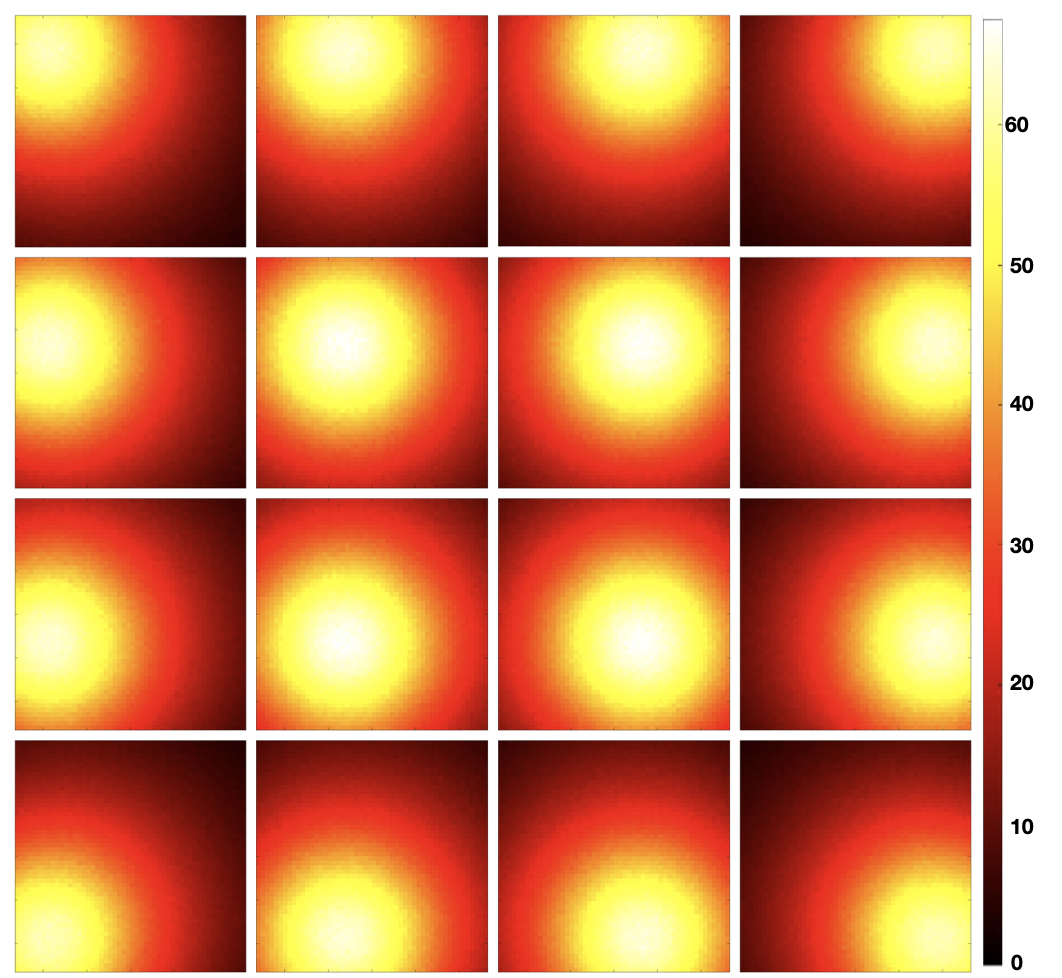}
			\includegraphics[scale=0.32]{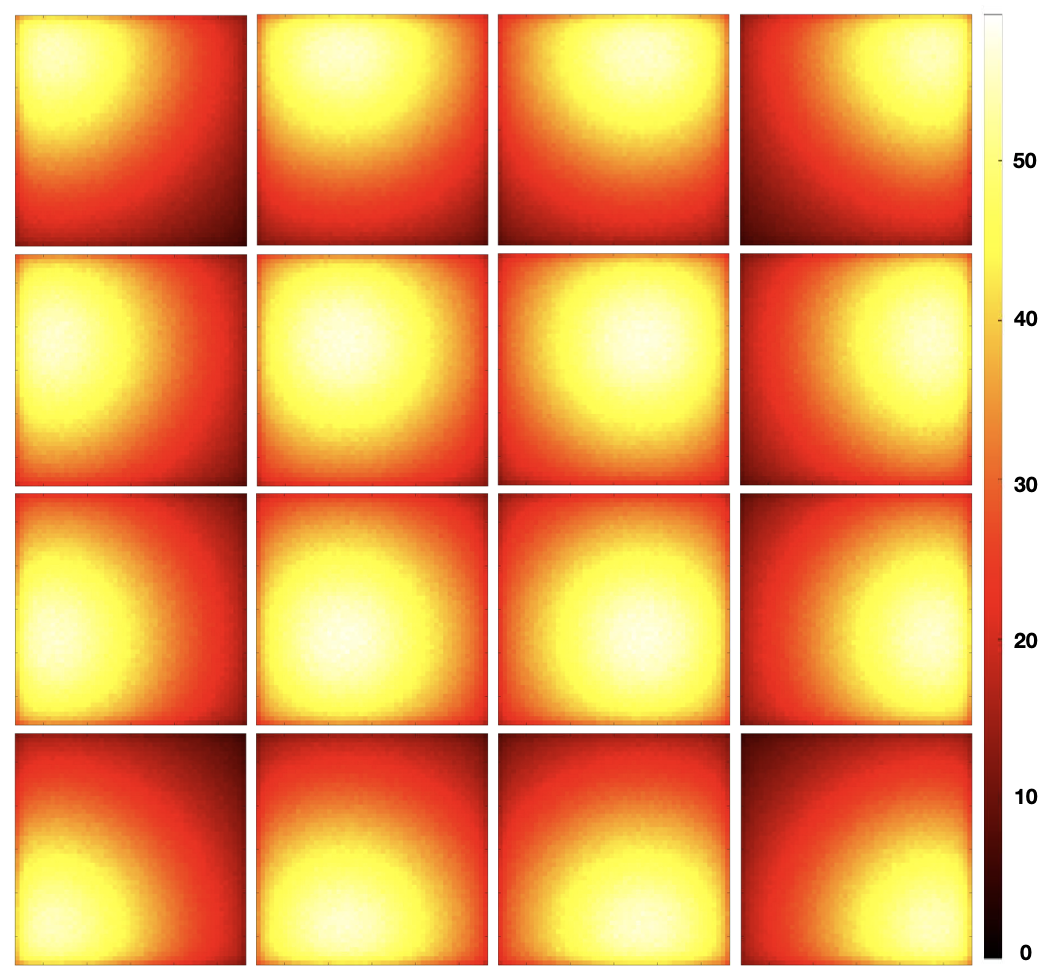}
			\caption{MDRFs (mean detector response functions, each map represents the mean signal response of a particular light sensor of the detector, as function of the interaction position) of simulated thick LYSO detector of section II, part C. The side length of each map is $26.0$mm. The MDRFs are achieved by scanning a visible-light point source in the scintillation crystal. The top figure shows the set of MDRFs located at $19.5$mm from the entrance surface of LYSO crystal (close to SiPM array). The middle figure shows the set of MDRFs on a plane located at a depth of $10.0$mm from the entrance surface of LYSO crystal. The bottom figure shows the set of MDRFs located at $0.5$mm from the entrance surface of LYSO crystal (close to detector entrance).}
		\end{center}
	\end{figure}\par

	Using the above sets of MDRFs to estimate the positions of events in the test dataset (the same dataset generated by scanning the scintillator with point light source) and comparing the estimated interaction position and the real interaction position of each event, the bias, standard deviation and root mean square error maps (in x, y and z directions) are shown in Fig. 10.

	\begin{figure} [H]
		\begin{center}
			\includegraphics[scale=0.3]{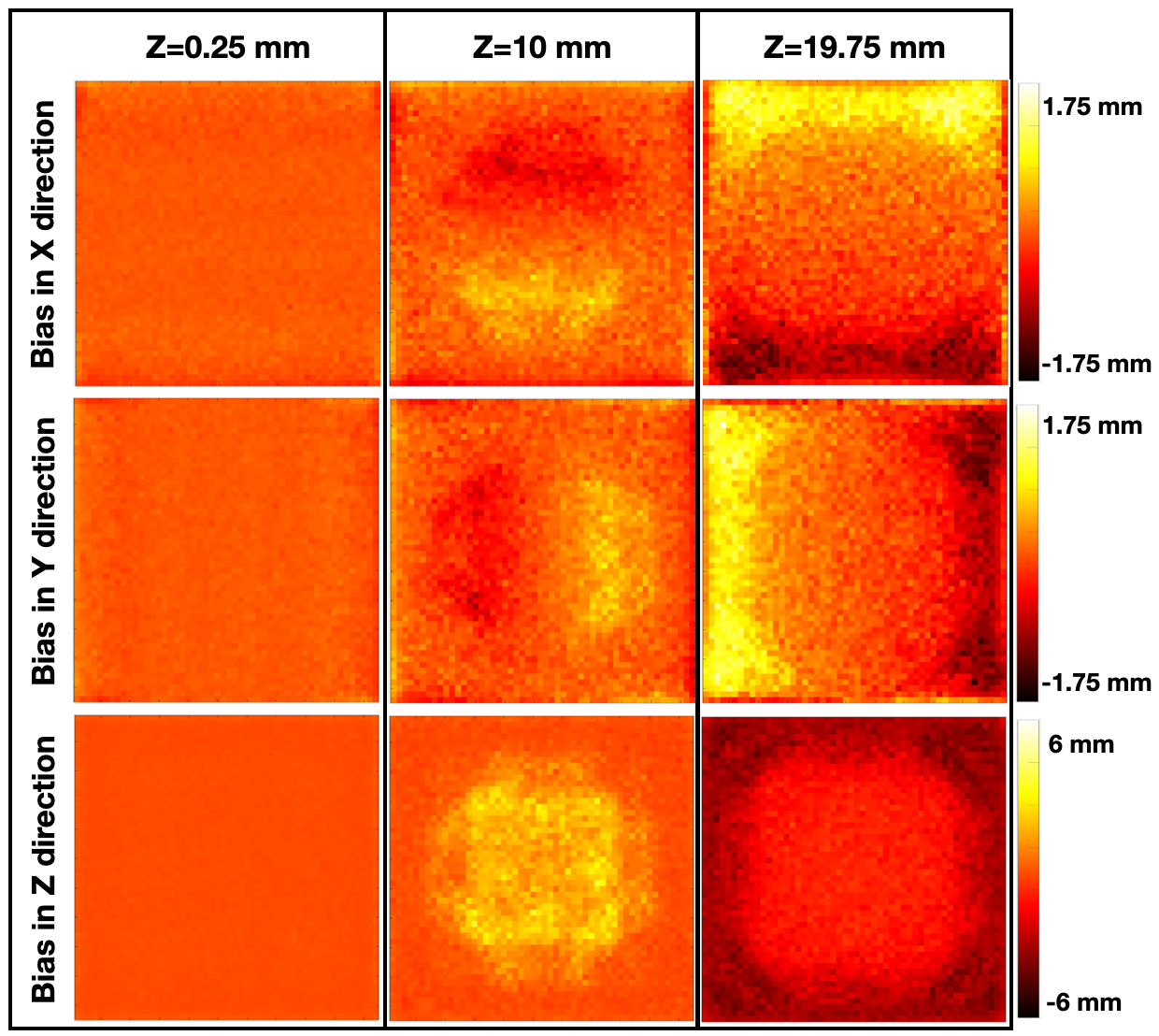}
			\includegraphics[scale=0.3]{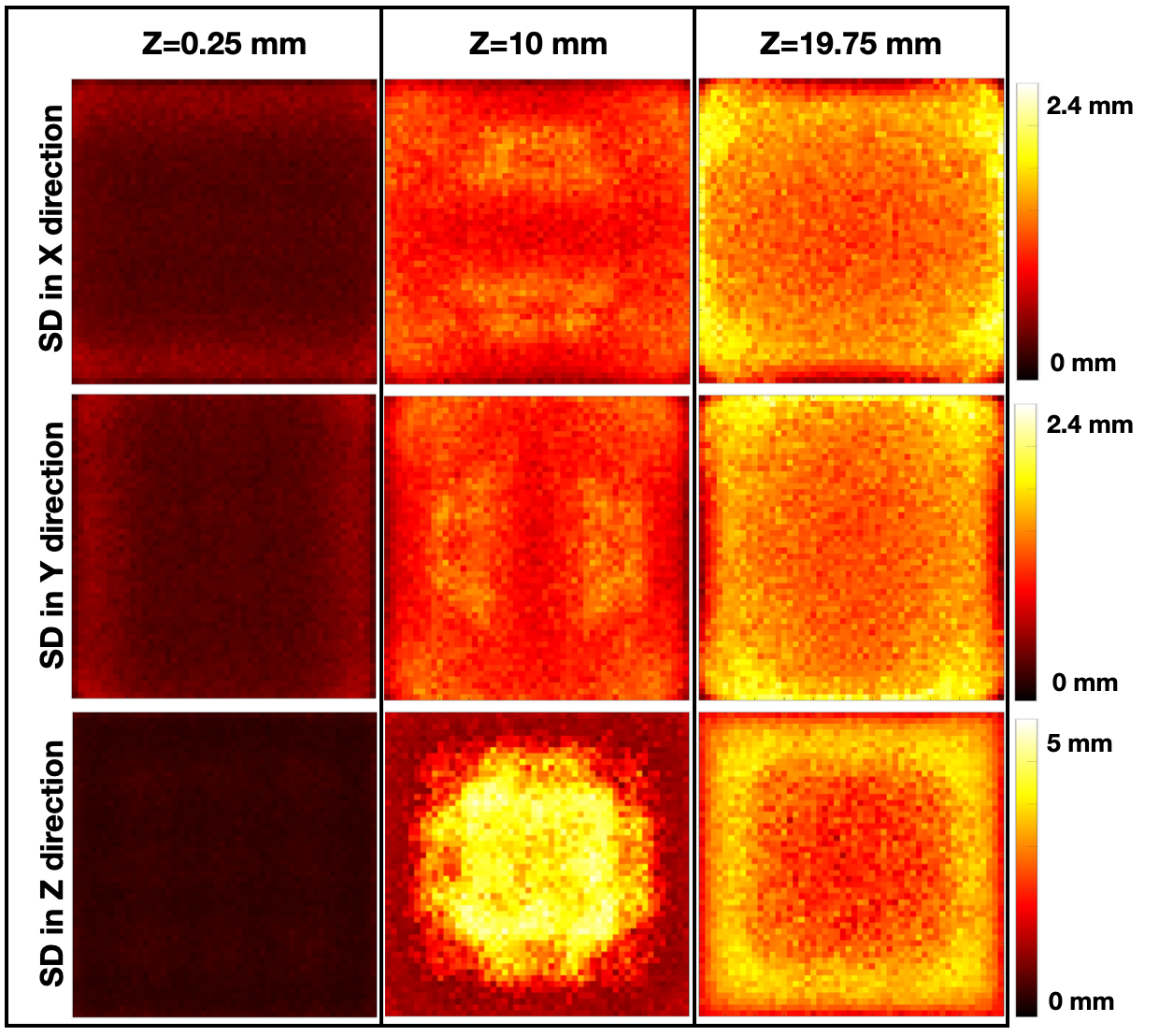}
			\includegraphics[scale=0.3]{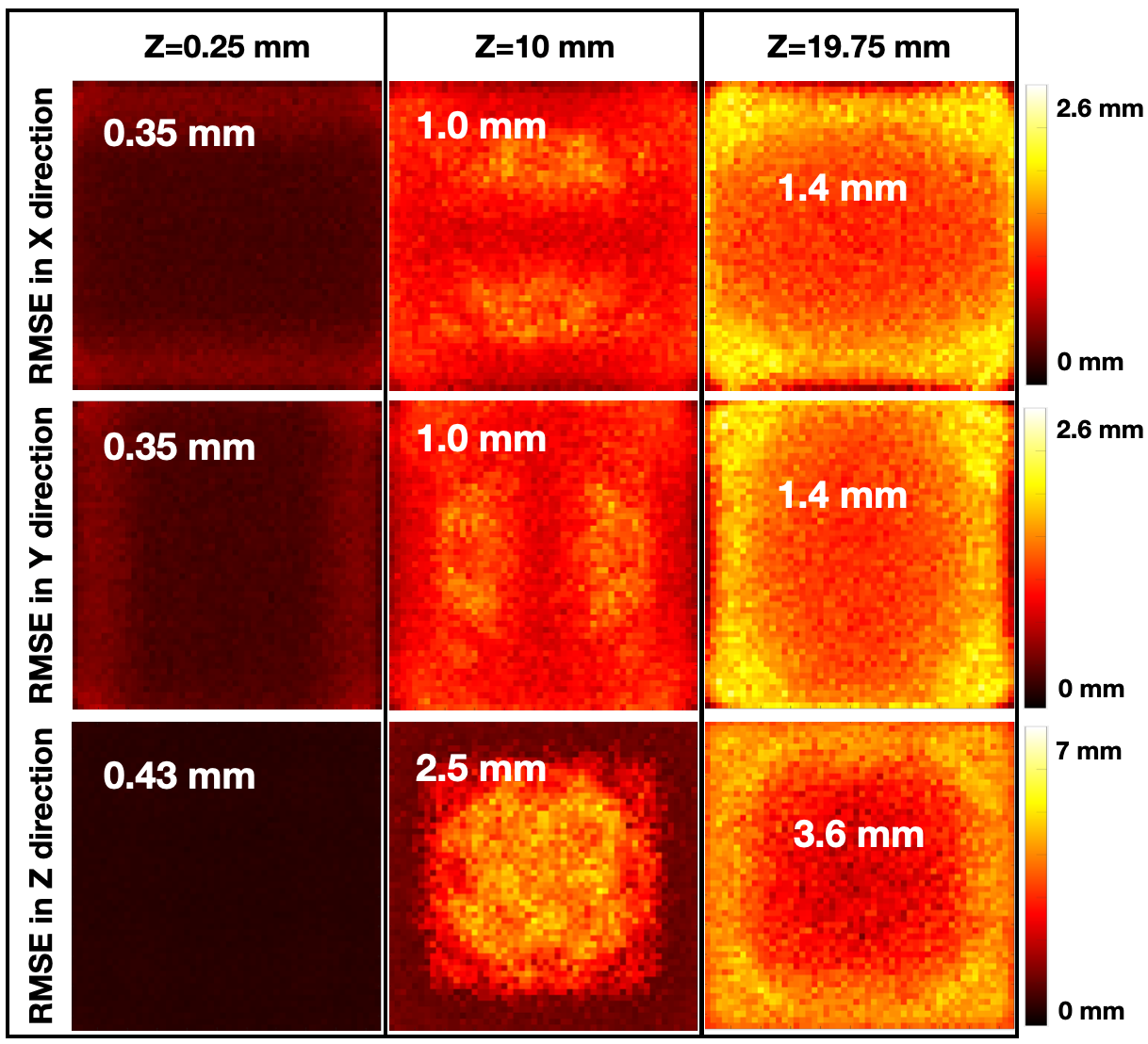}
			\caption{The bias (top), standard deviation (middle) and root mean square error (bottom) maps of estimated interaction positions using the MDRFs in Fig. 9 (ground truth MDRFs). In root men square error maps, the mean for each map is shown as text on top of each map. Each map's dimension is $26$mm$\times26$mm.}
		\end{center}
	\end{figure}\par
	
	\subsubsection{Simulation excluding Compton effect}
	The CDS-generated MDRFs without considering Compton effect of the thick scintillation crystal detector of section II, part C(1) is shown in Fig. 11. Again, only three representative sets of MDRFs at three different depths are shown.\par
	
	\begin{figure} [H]
		\begin{center}
			\includegraphics[scale=0.32]{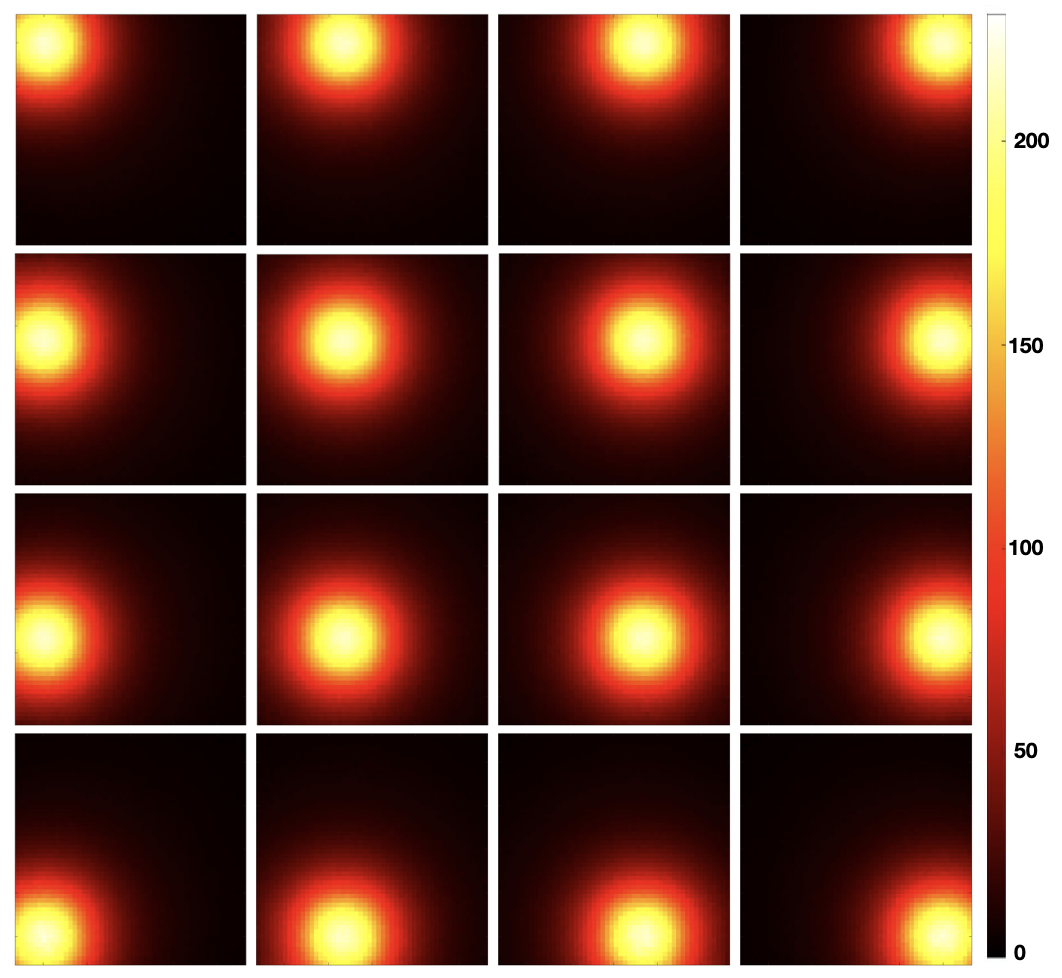}
			\includegraphics[scale=0.32]{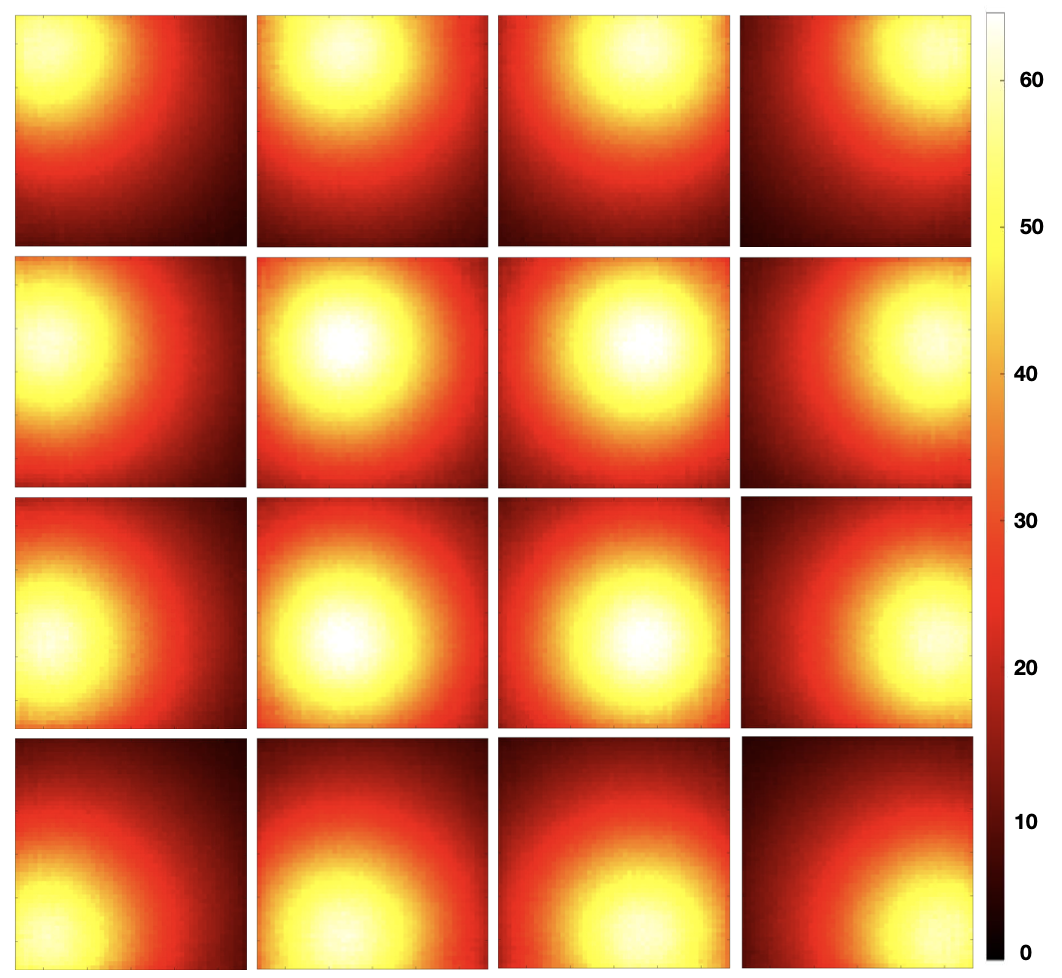}
			\includegraphics[scale=0.32]{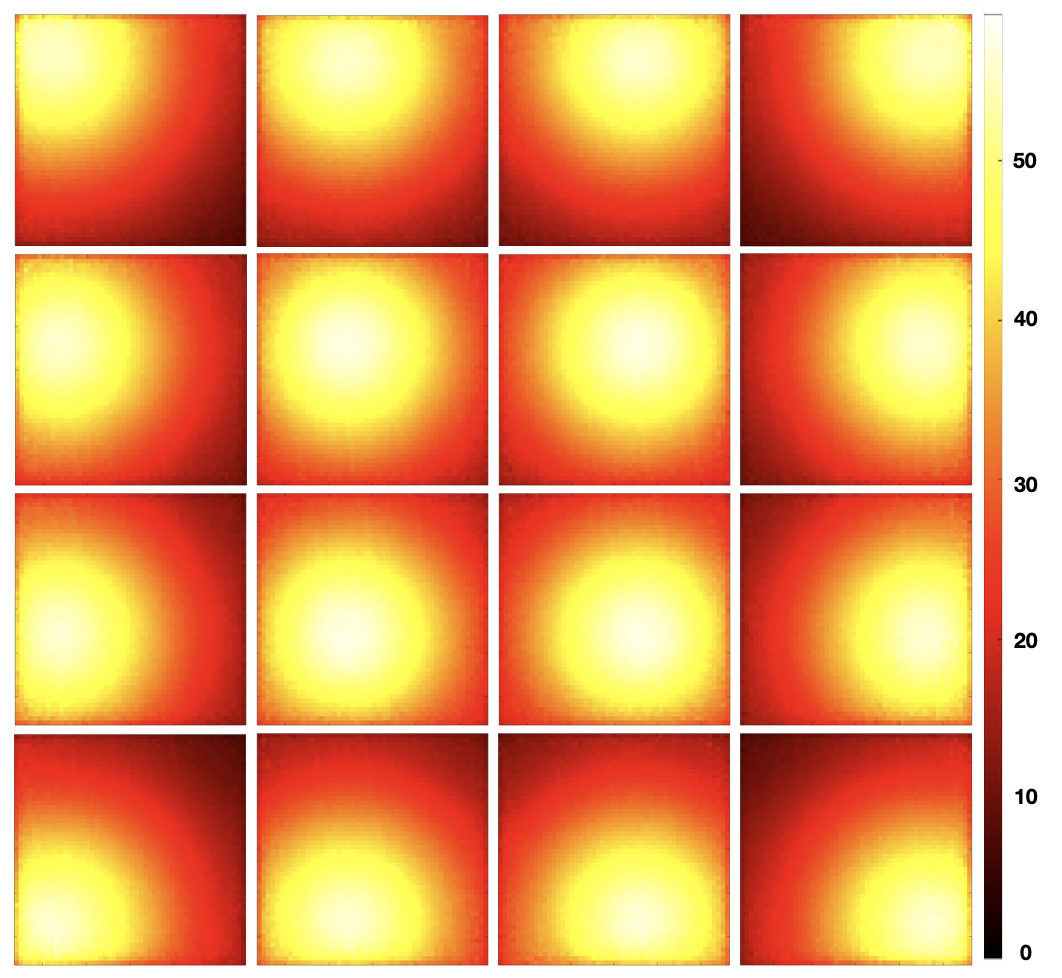}
			\caption{The MDRFs are achieved using the CDS method, using the datasets without Compton effect. The top figure shows the MDRFs located at a depth of $19.5$mm from the crystal entrance surface. The middle figure shows the MDRFs located at a depth of $10.0$mm from the entrance surface. The bottom figure shows the MDRFs located at $0.5$mm from the entrance surface.}
		\end{center}
	\end{figure}\par
	
	If we subtract Fig. 9 from Fig. 11, and calculate the difference relative to the maximum value in the ground truth MDRF (Fig. 9), an error map can be created, as shown in Fig. 12, the color bar is in logarithmic scale.\par
	
	\begin{figure} [H]
		\begin{center}
			\includegraphics[scale=0.3]{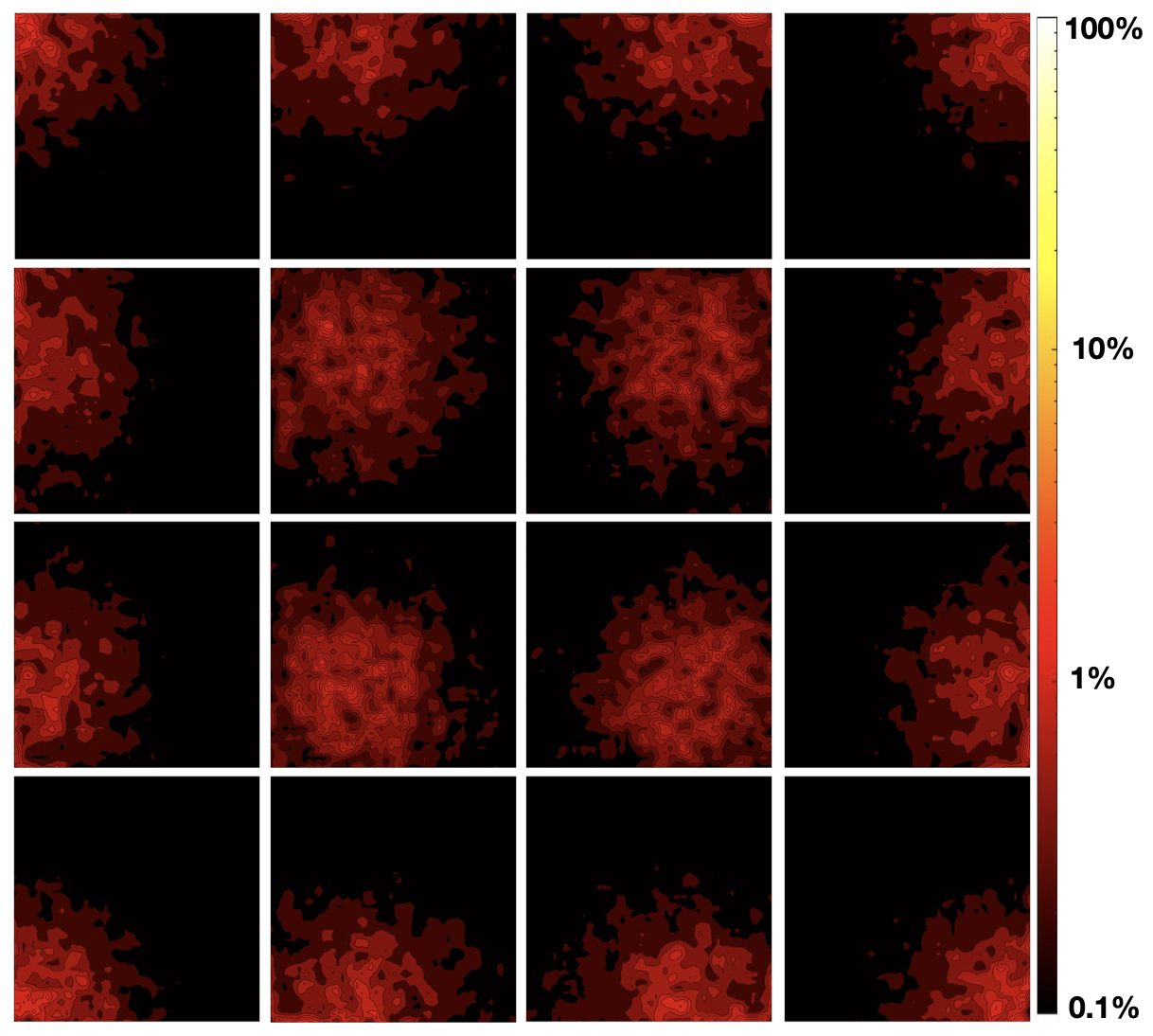}
			\includegraphics[scale=0.3]{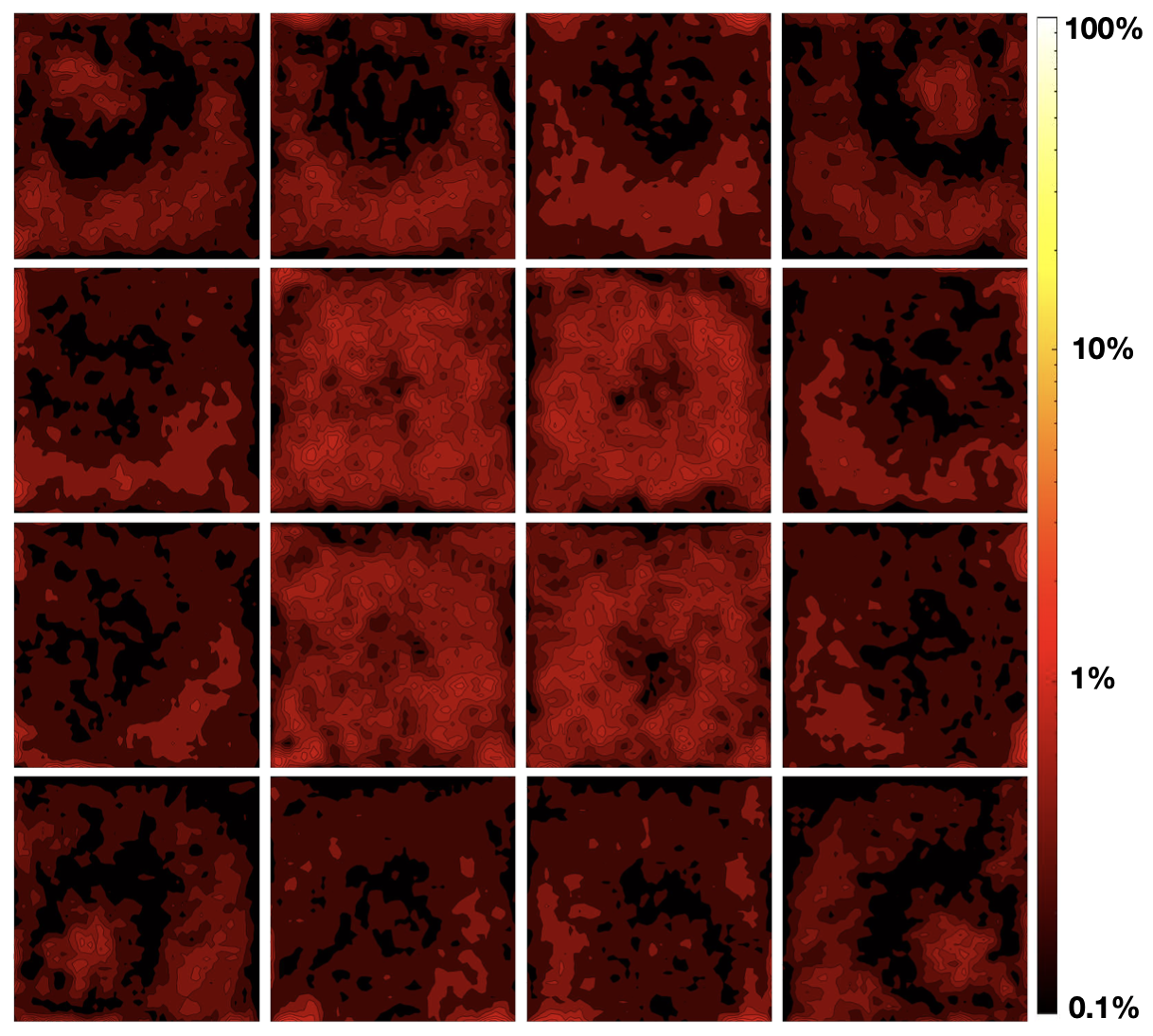}
			\includegraphics[scale=0.3]{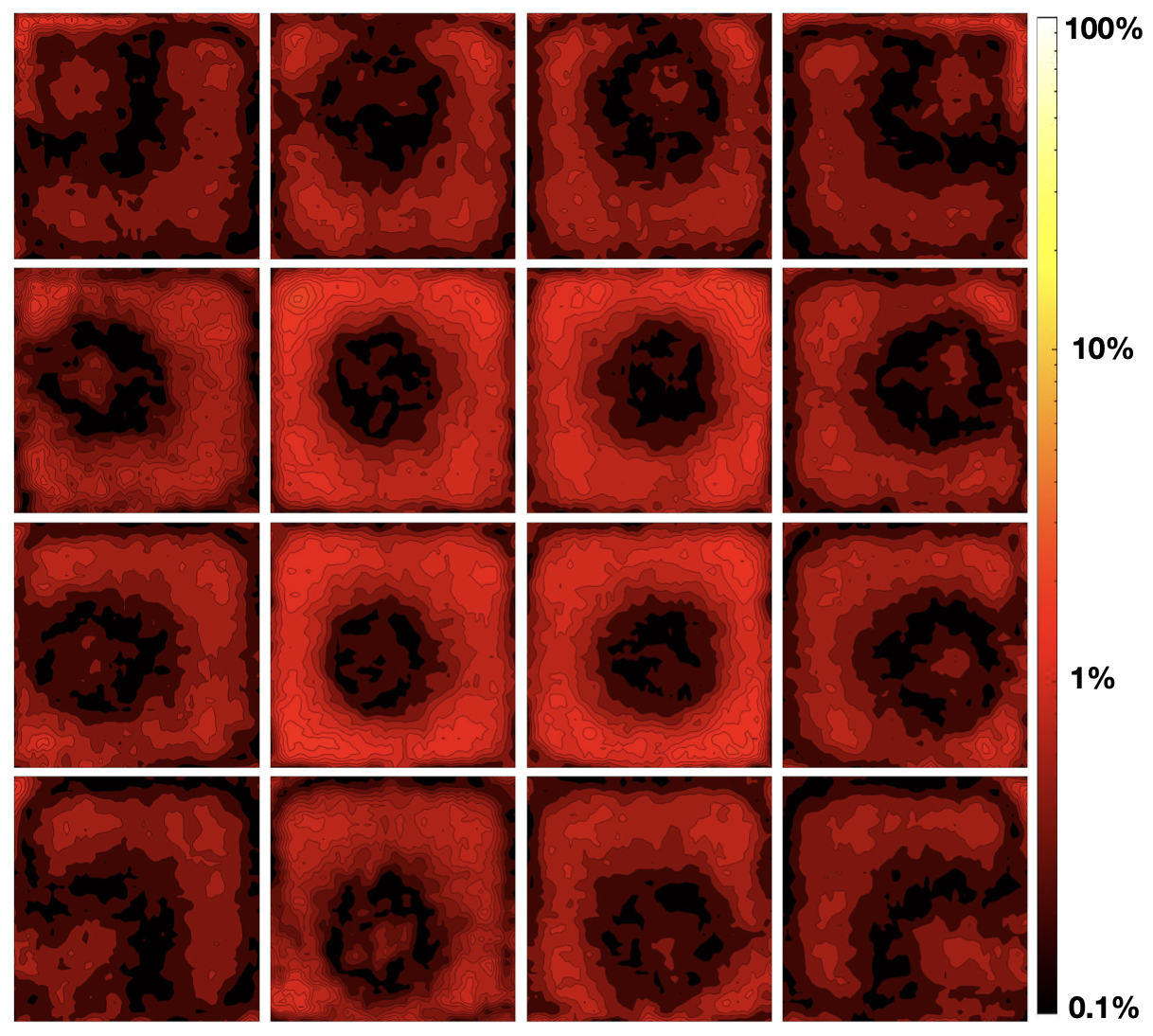}
			\caption{The corresponding logarithmic error maps by comparing Fig.9 and Fig. 11, the corresponding mean and maximum error of each map is shown in TABLE I. The color bar was set to be the percentage of maximum MDRF value of Fig. 9.}
		\end{center}
	\end{figure}\par
	
	The mean and maximum errors of each error map are listed in TABLE I below.
	
	\begin{table}[H]
		\fontsize{5}{7}\selectfont
		\centering
		\caption{Mean and maximum values of each error map of Fig. 12, as percentage of maximum value of Fig. 9.}
		\label{T:peak}
		\includegraphics[scale=0.6]{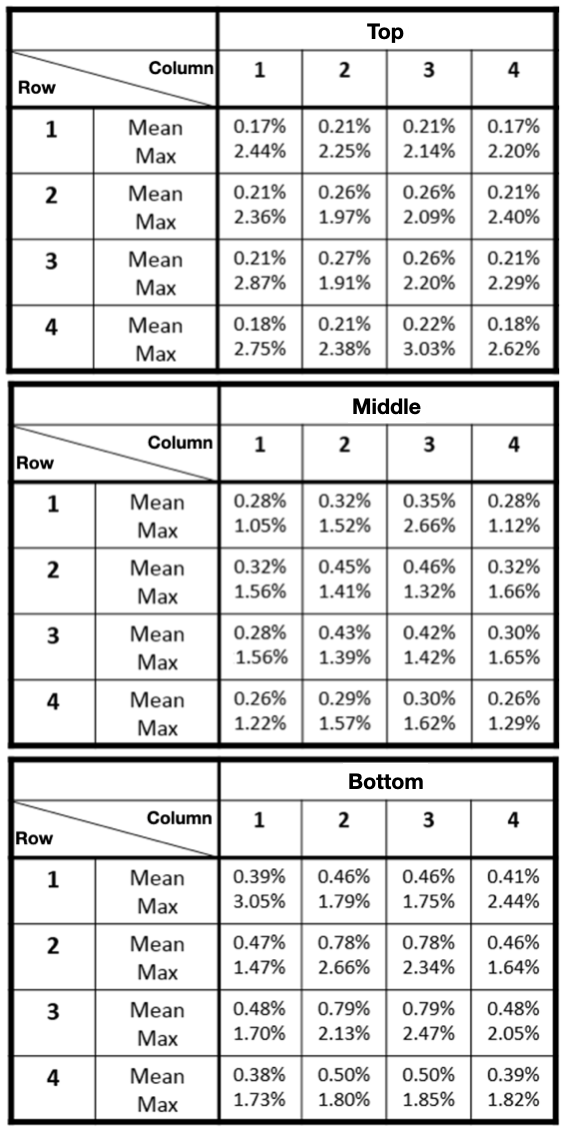}	
	\end{table}

	Using the MDRFs in Fig. 11, the positions of the test events were estimated and compared with their corresponding true positions. The bias, standard deviation and root mean square error maps are shown in Fig. 13.

	\begin{figure} [H]
		\begin{center}
			\includegraphics[scale=0.3]{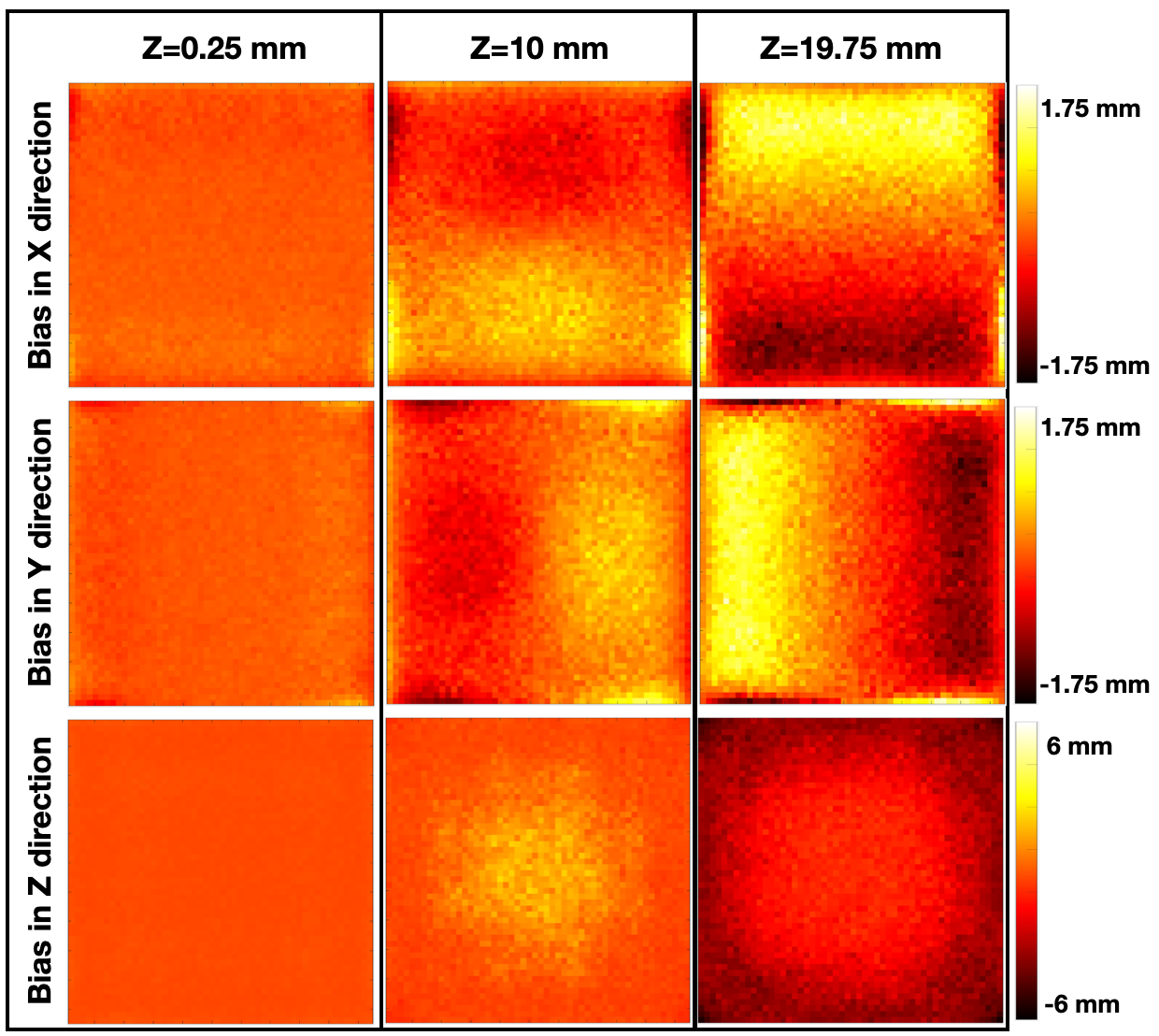}
			\includegraphics[scale=0.3]{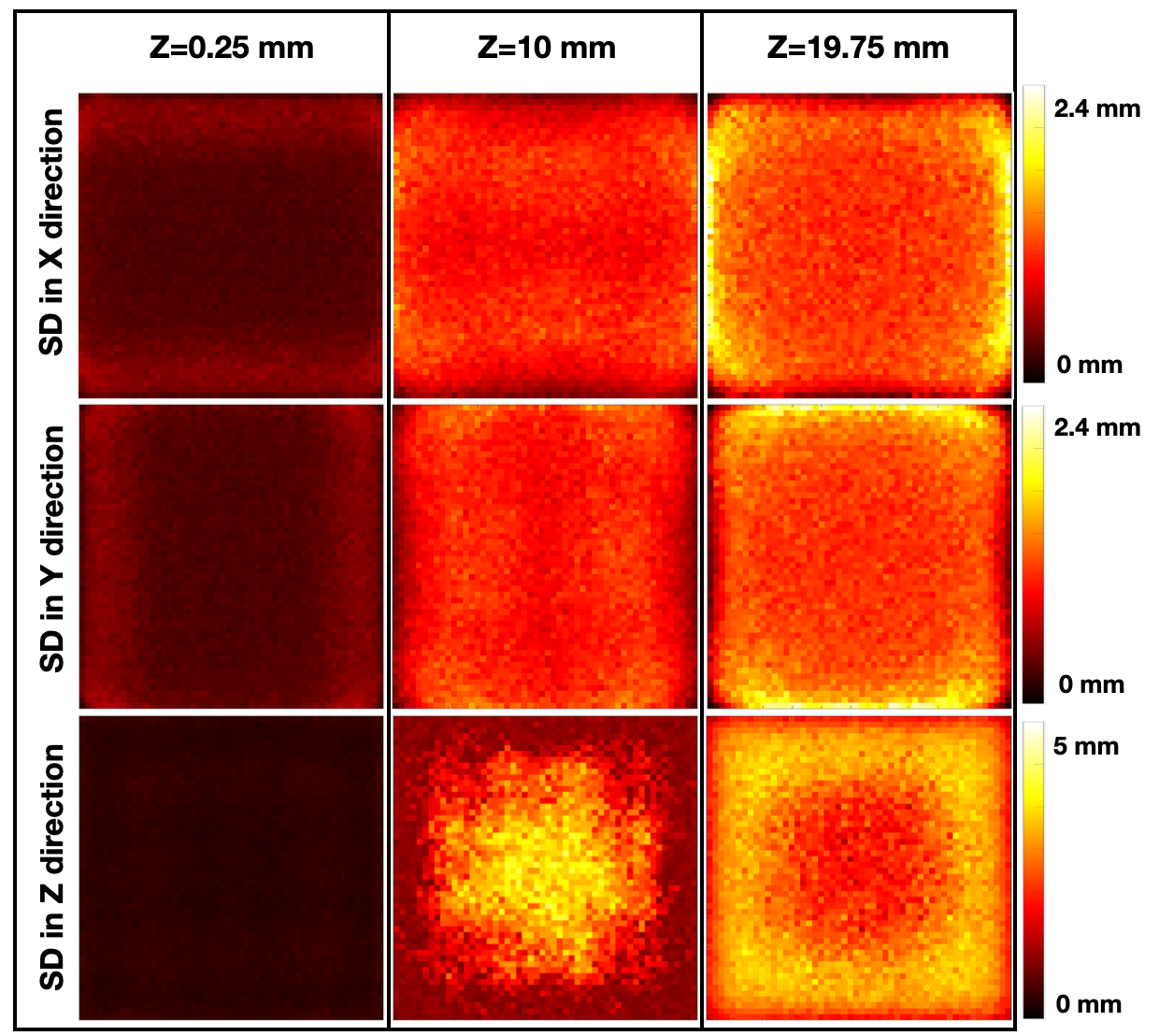}
			\includegraphics[scale=0.3]{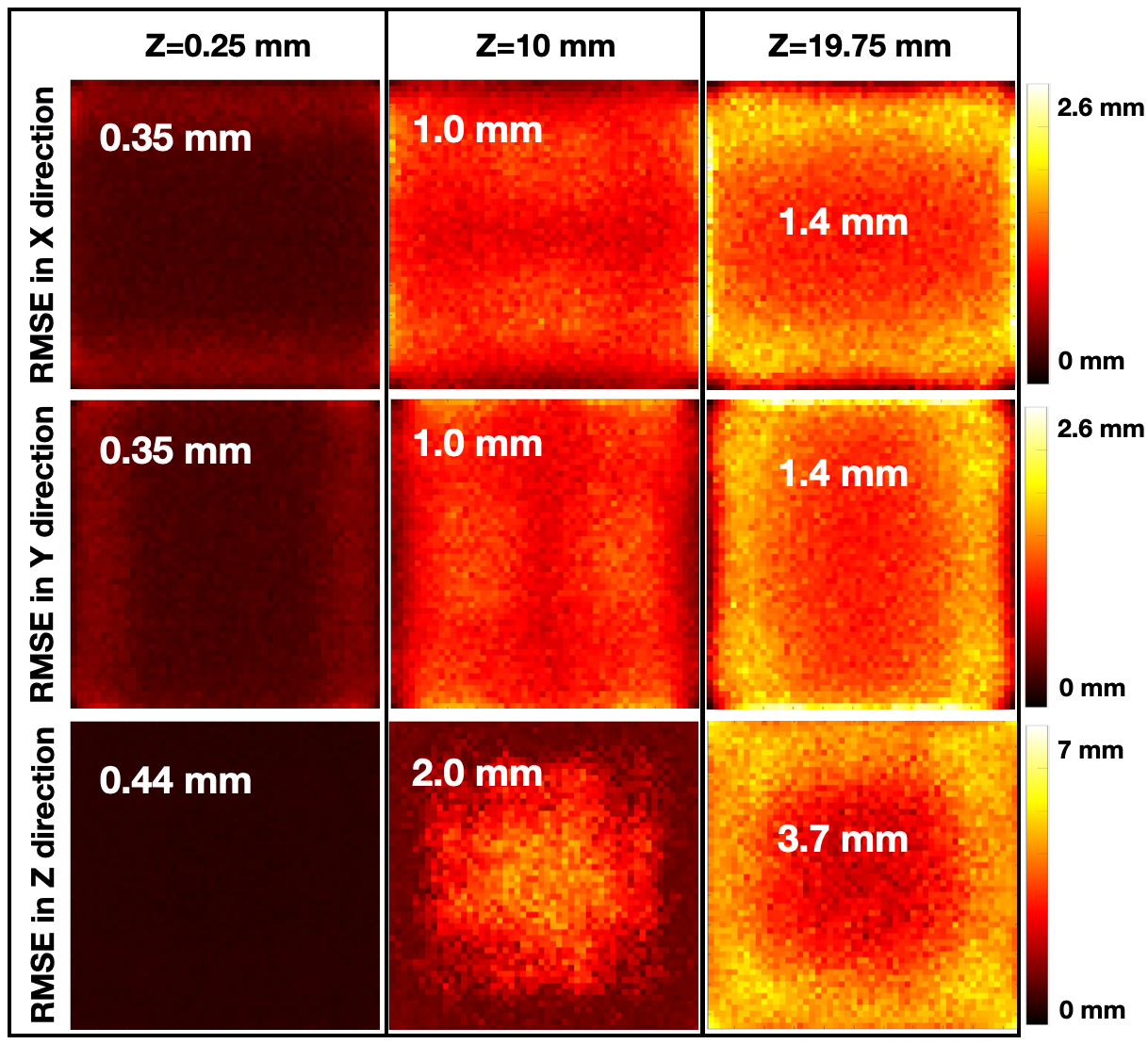}
			\caption{The bias (top), standard deviation (middle) and root mean square error (bottom) maps of estimated interaction positions using the MDRFs in Fig. 11 (MDRFs generated with CDS method without considering Compton effect). In root mean square error maps, the mean for each map is shown as text on top of each map.}
		\end{center}
	\end{figure}\par
	
	\subsubsection{Simulation including Compton effect}
	
	Due to the fact that Compton effect can cause gamma-ray photon energy to be deposited partially, these partial-energy events need to be filtered out. Example of a histogram showing distribution of detected visible photon number is presented in Fig. 14. The events making this histogram are from a dataset of the fan beam scanning the detector center along x direction. There are in total $500,000$ events in this dataset for better illustration. An energy window of $247-824$ keV, which corresponds to $300-1,000$ detected photons in Fig. 14, was used to filter out the partial energy events.\par
	
	\begin{figure} [H]
		\begin{center}
			\includegraphics[scale=0.35]{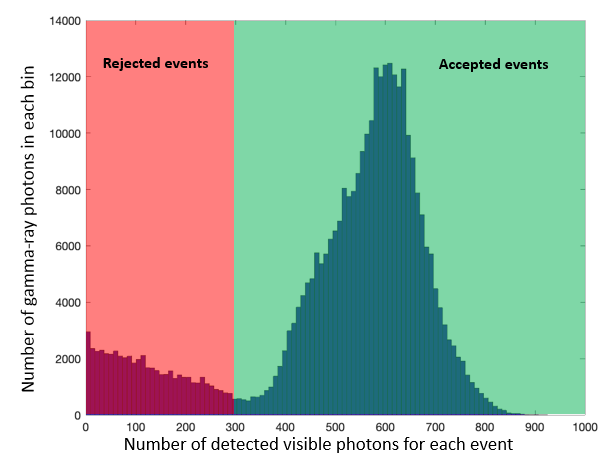}
			\caption{A histogram showing the distribution of gamma-ray interactions by the number of detected visible photons. There are in total $50,000$ events in the histogram, and these events were achieved by a fan beam scanning along x direction at the center of the detector.}
		\end{center}
	\end{figure}\par
	
	After filtering out the partial-energy events, the datasets were used to calculate the MDRFs using the CDS method. The MDRFs achieved are shown in Fig. 15.\par
	
	\begin{figure} [H]
		\begin{center}
			\includegraphics[scale=0.26]{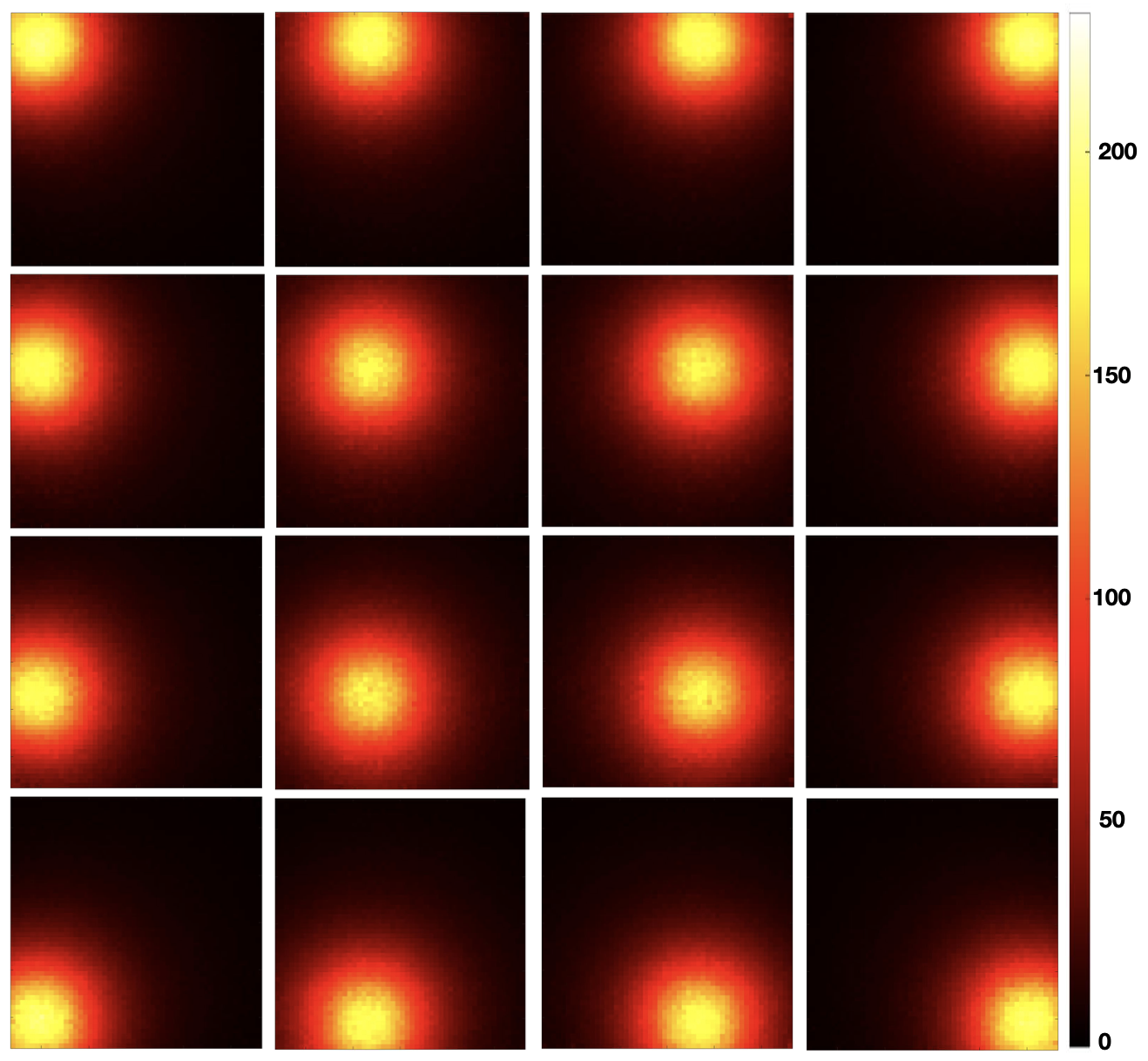}
			\includegraphics[scale=0.26]{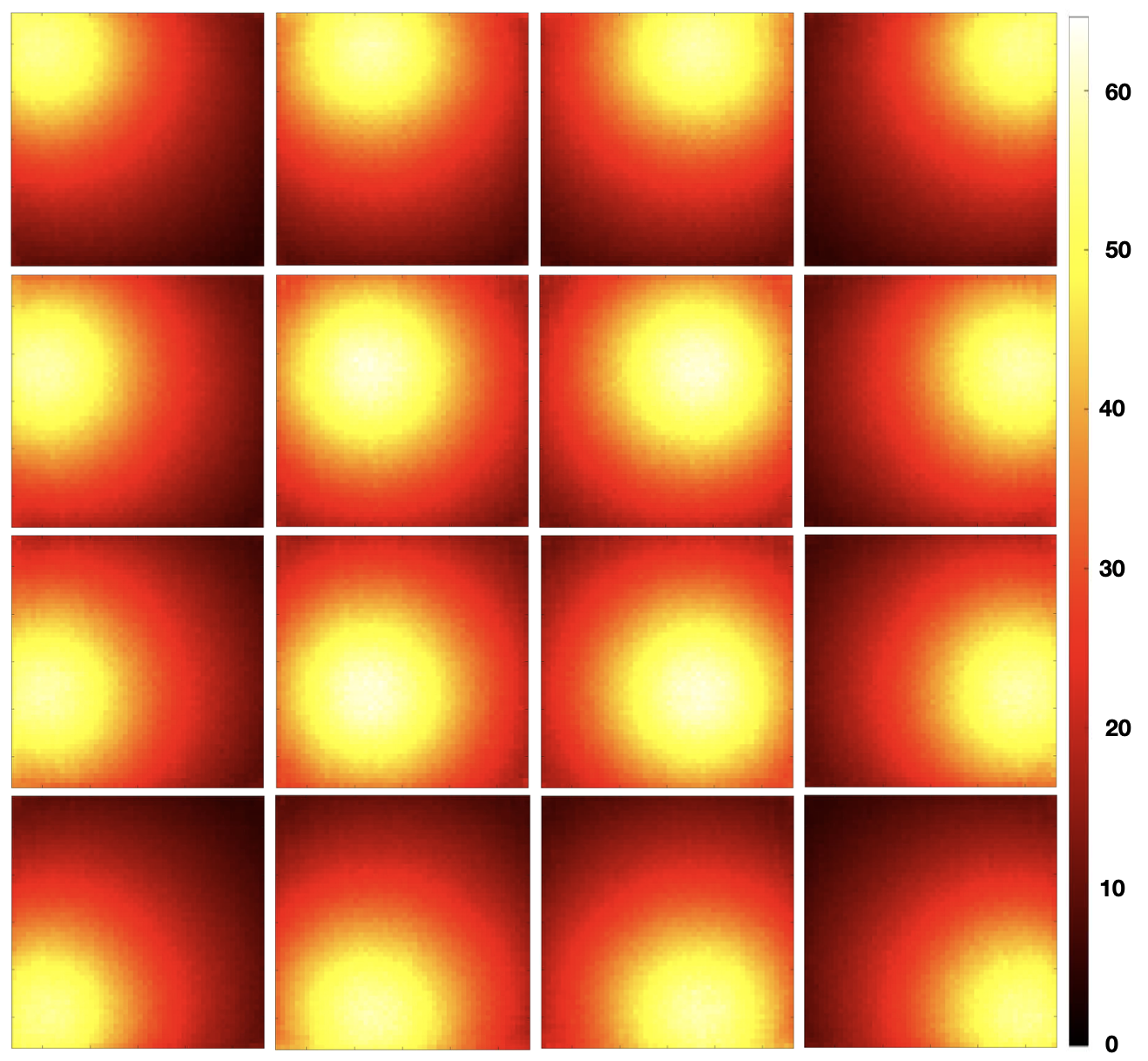}
			\includegraphics[scale=0.26]{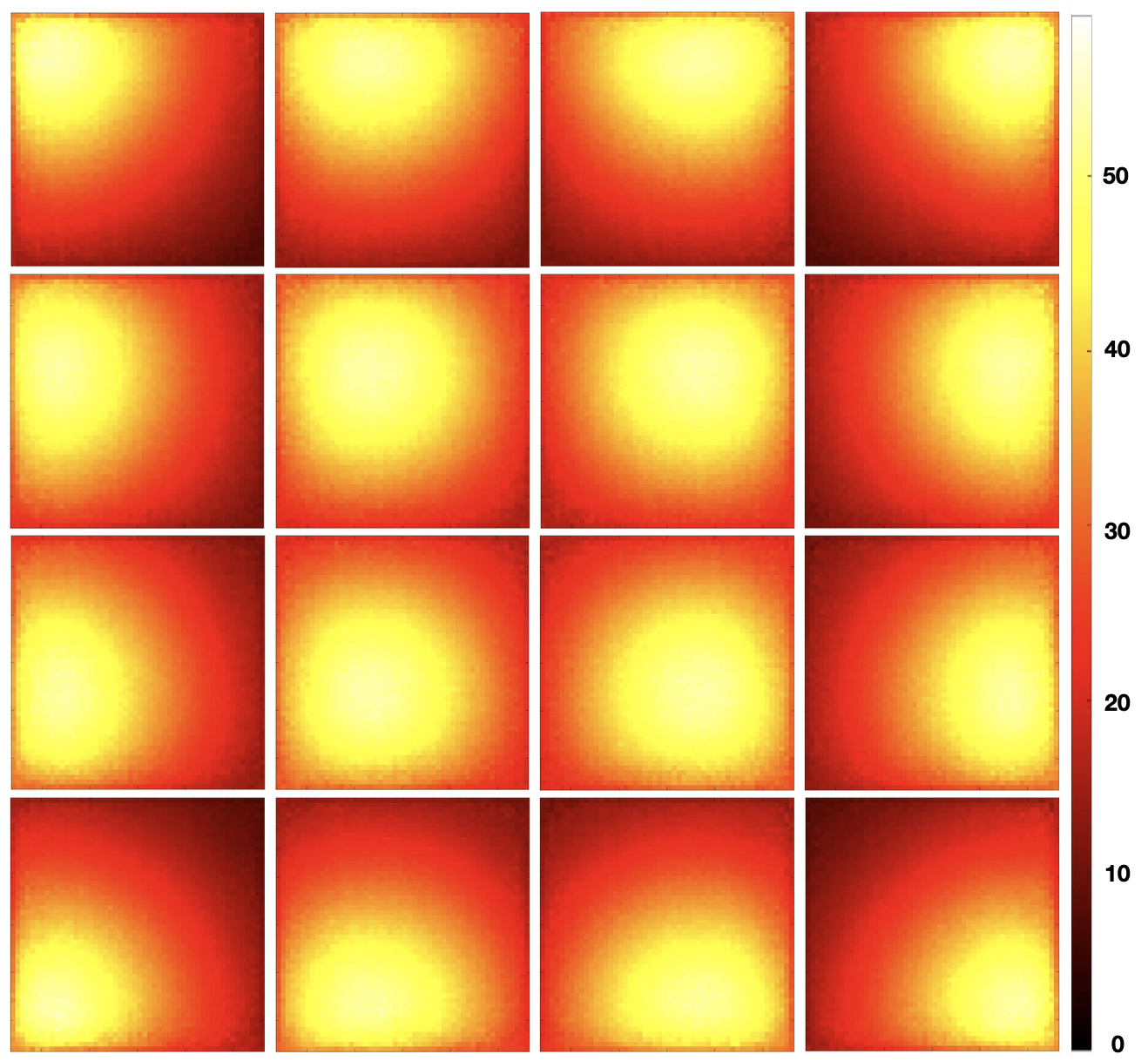}
			\caption{The MDRFs are achieved using the CDS method considering Compton effect. The top figure shows the MDRFs located at a depth of $19.5$mm from the crystal entrance surface. The middle figure shows the MDRFs located at a depth of $10.0$mm from the entrance surface. The bottom figure shows the MDRFs located at $0.5$mm from the entrance surface.}
		\end{center}
	\end{figure}\par
	
	By comparing Fig. 15 with Fig. 9, the error maps can be calculated. Those error maps are shown in Fig. 16.\par
	
	\begin{figure} [H]
		\begin{center}
			\includegraphics[scale=0.23]{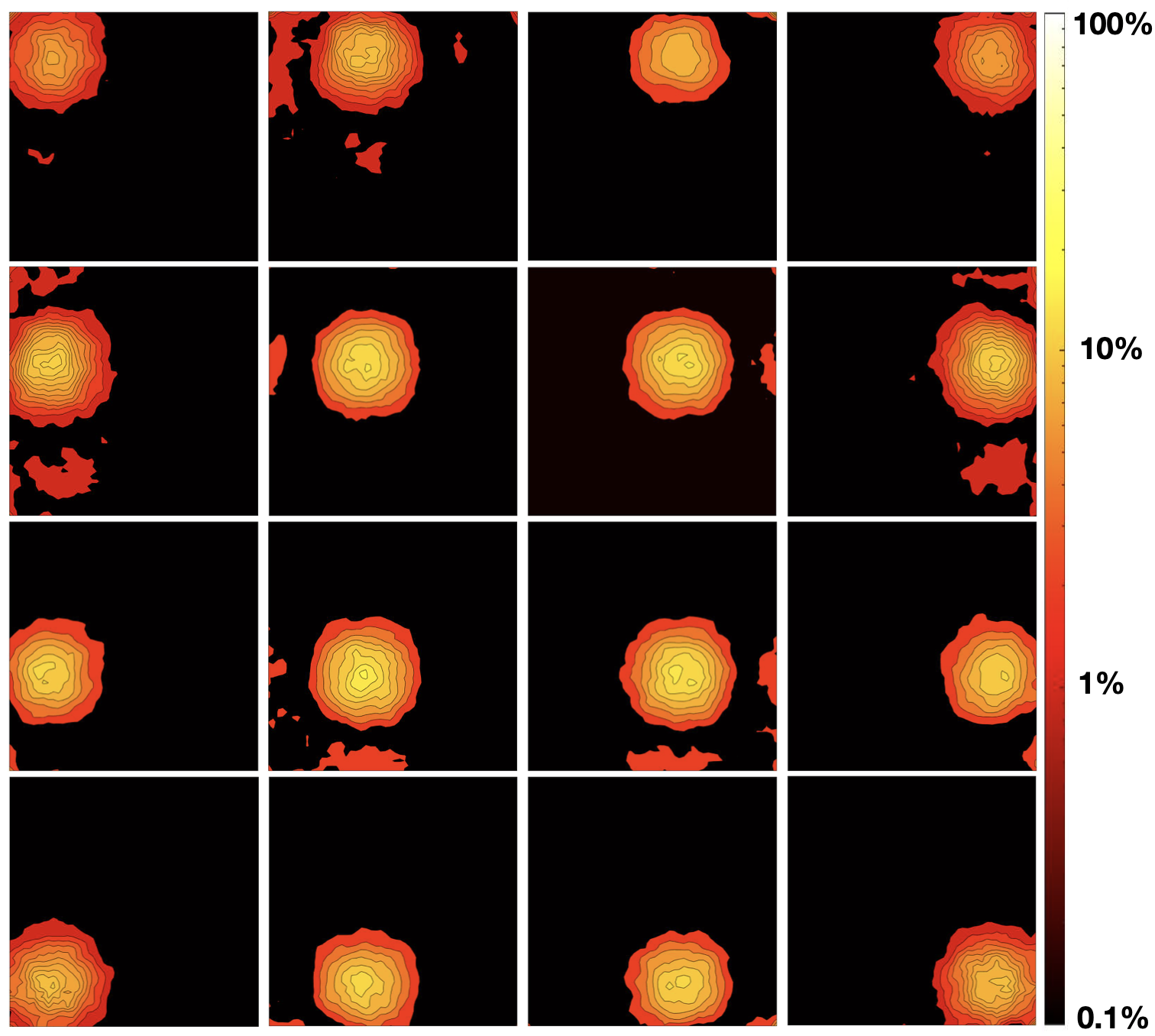}
			\includegraphics[scale=0.23]{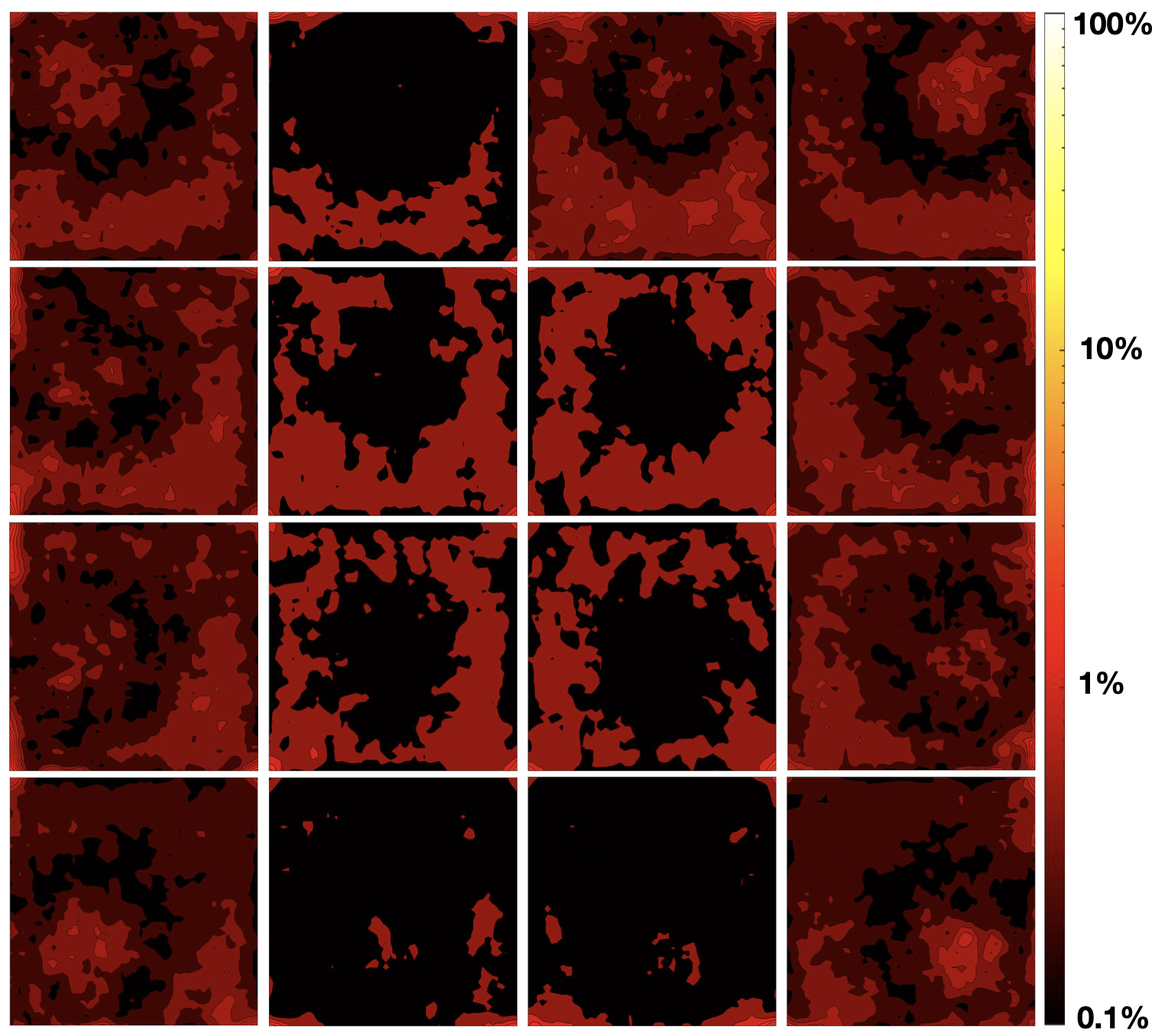}
			\includegraphics[scale=0.23]{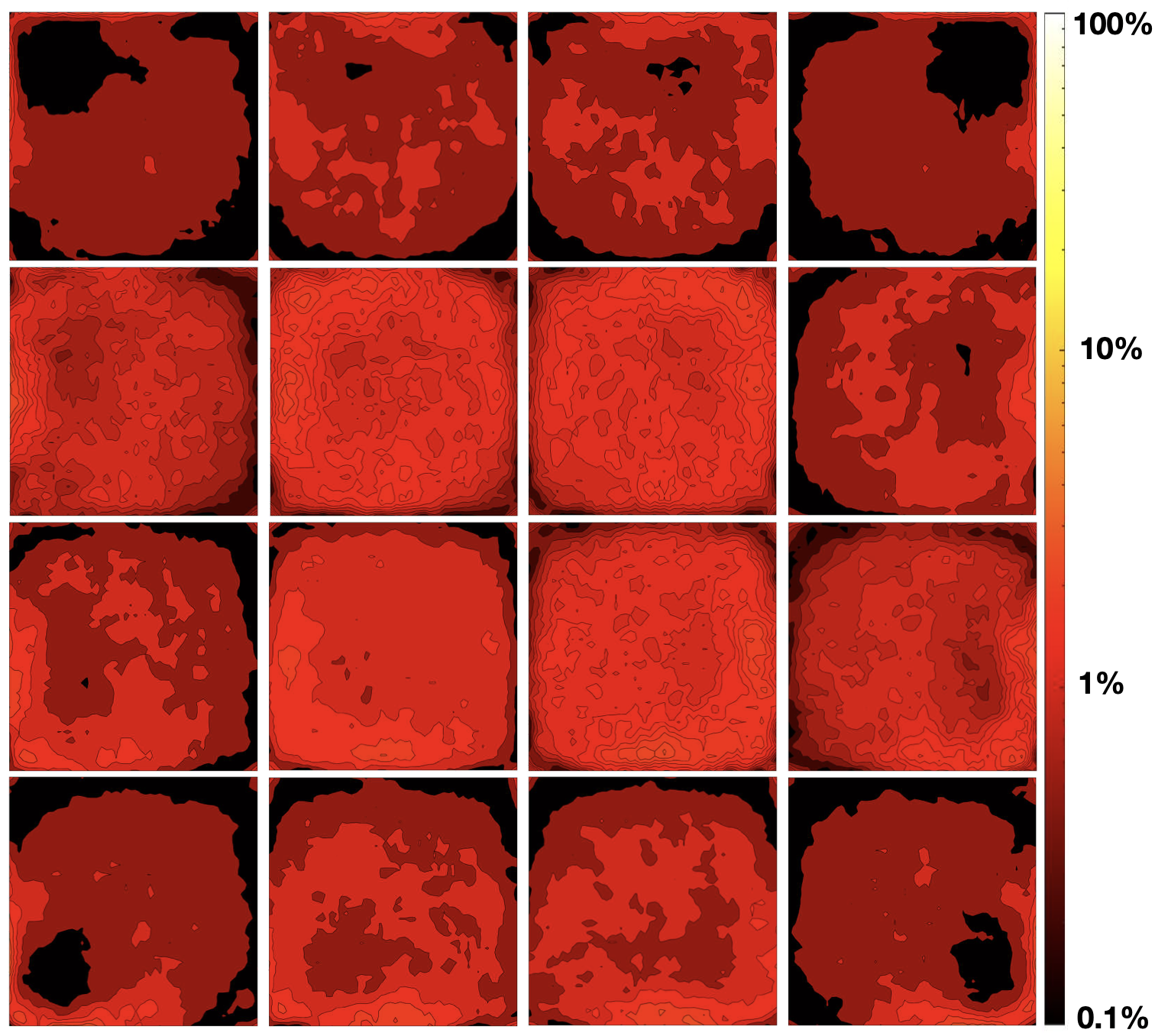}
			\caption{The corresponding logarithmic error maps by comparing Fig. 15 and Fig. 9, the corresponding mean and maximum error of each map is shown in TABLE II. The color bar was set to be the percentage of maximum MDRF value of Fig. 9. The relatively large error in the upper maps is majorly due to the Compton effect (by blurring out the MDRFs with multi-scatter events).}
		\end{center}
	\end{figure}\par
	
	\begin{table}[H]
		\fontsize{5}{7}\selectfont
		\centering
		\caption{Mean and maximum value of each error map of Fig. 16, as percentage of maximum map value of Fig. 9. The orders of the maps are preserved.}
		\label{T:peak}
		\includegraphics[scale=0.6]{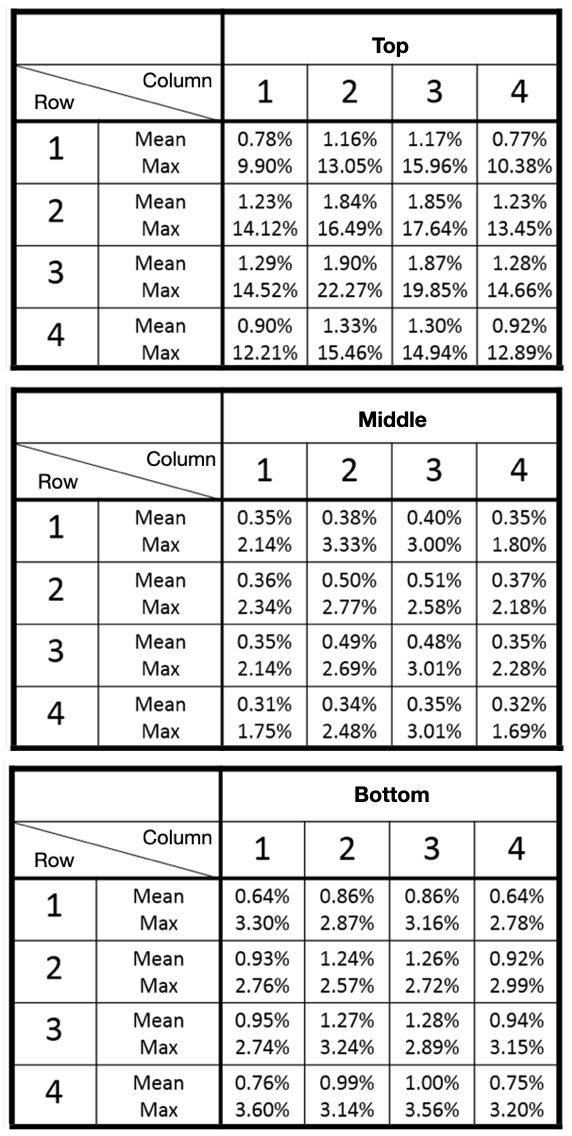}	
	\end{table}

	Using the MDRFs in Fig. 15, the positions of the test events were estimated and compared with their corresponding true positions. The bias, standard deviation and root mean square error maps are shown in Fig. 17.
	
	\begin{figure} [H]
		\begin{center}
			\includegraphics[scale=0.3]{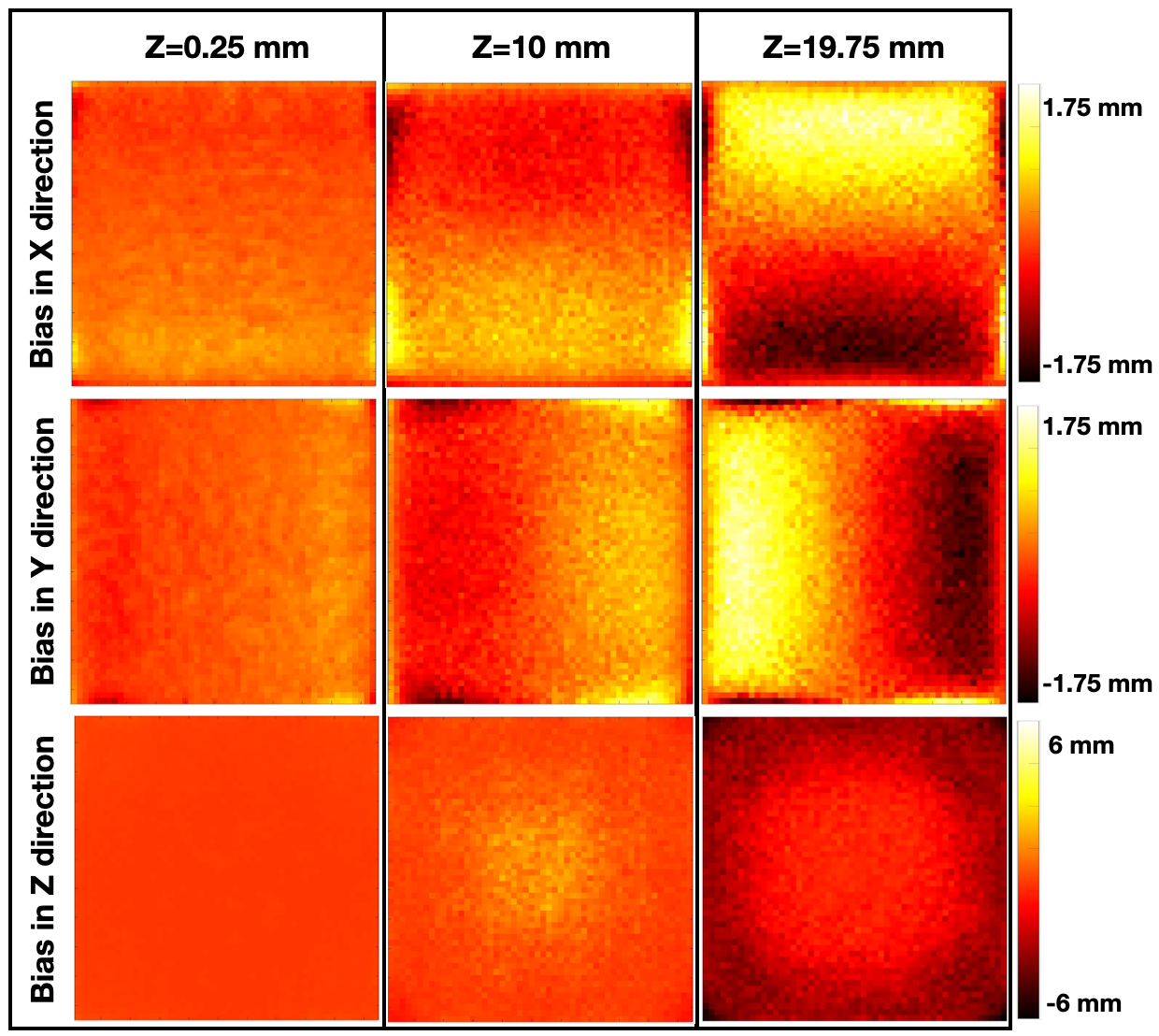}
			\includegraphics[scale=0.3]{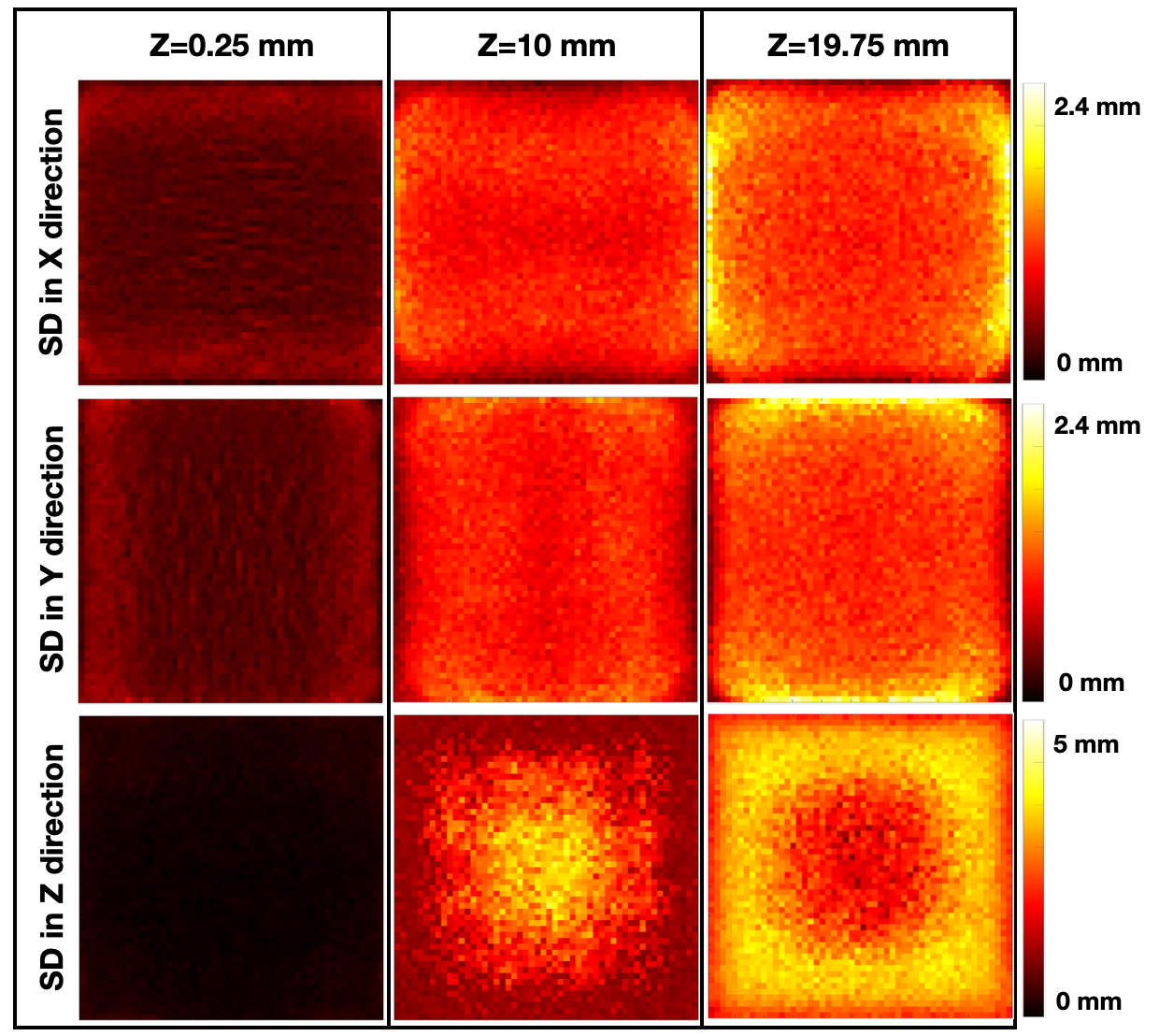}
			\includegraphics[scale=0.3]{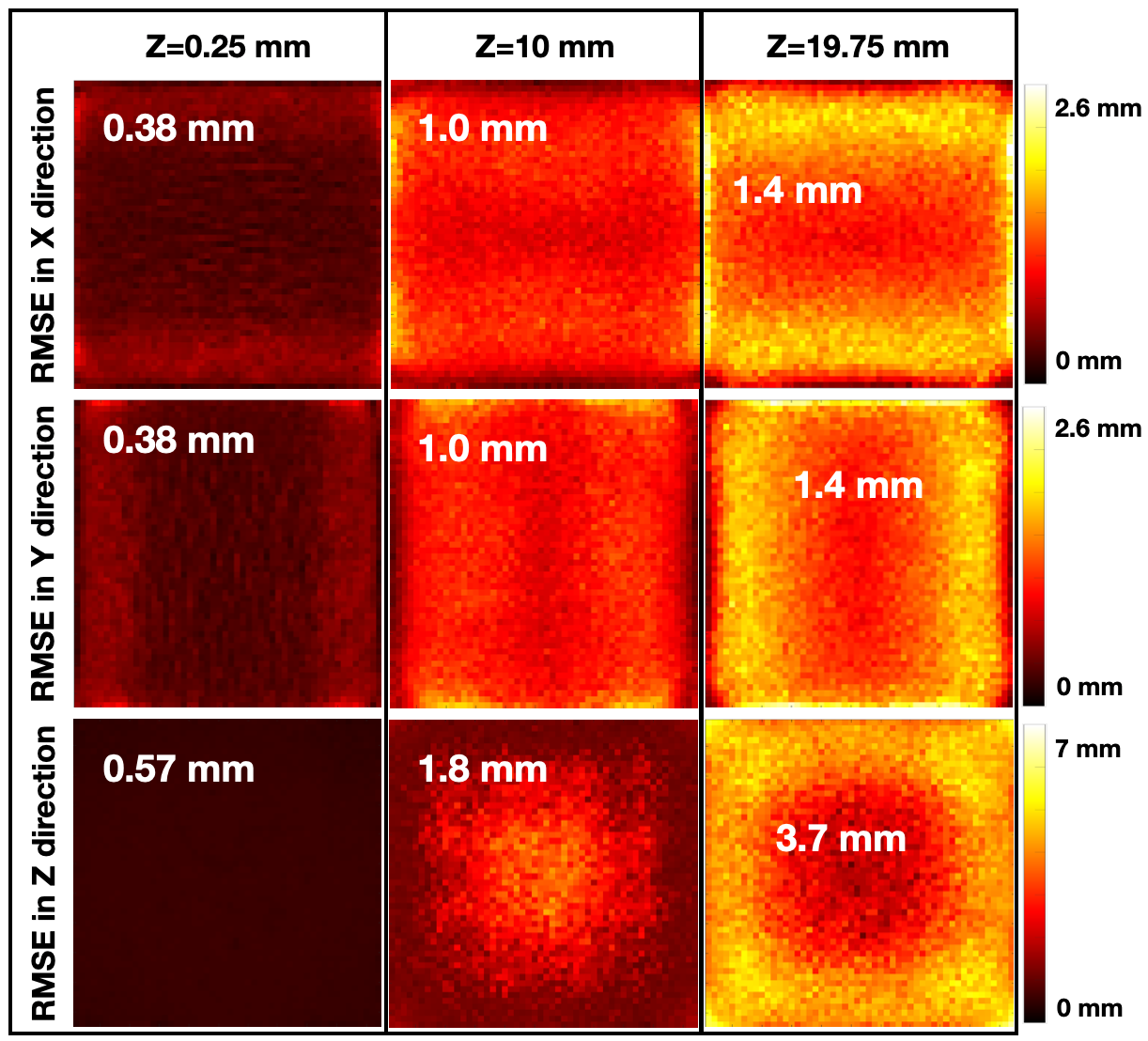}
			\caption{The bias (top), standard deviation (middle) and root mean square error (bottom) maps of estimated interaction positions using the MDRFs in Fig. 15 (MDRFs generated with CDS method considering Compton effect). In root mean square error maps, the mean error for each map is shown as text on top of each map.}
		\end{center}
	\end{figure}\par
	
	\subsection{Simulation of an edge-readout detector}
	The MDRFs generated using a traditional scanning pencil-beam method are shown in Fig. 18 (top). The MDRFs generated using the CDS method are shown in Fig. 18 (middle). The logarithmic error maps are shown in Fig. 18 (bottom), with its corresponding mean and maximum value shown in TABLE III.
	
	\begin{figure} [H]
		\begin{center}
			\includegraphics[scale=0.23]{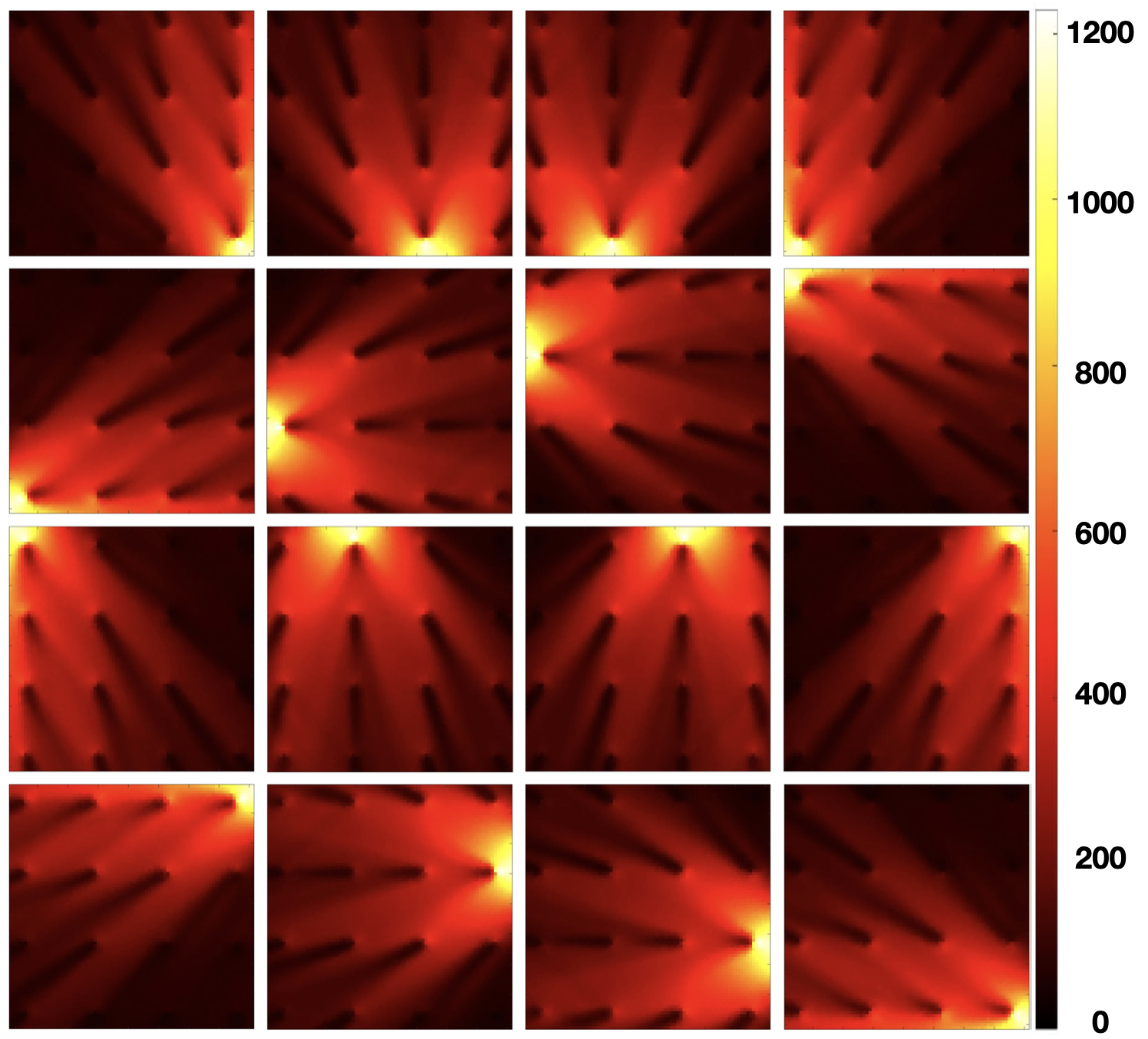}
			\includegraphics[scale=0.23]{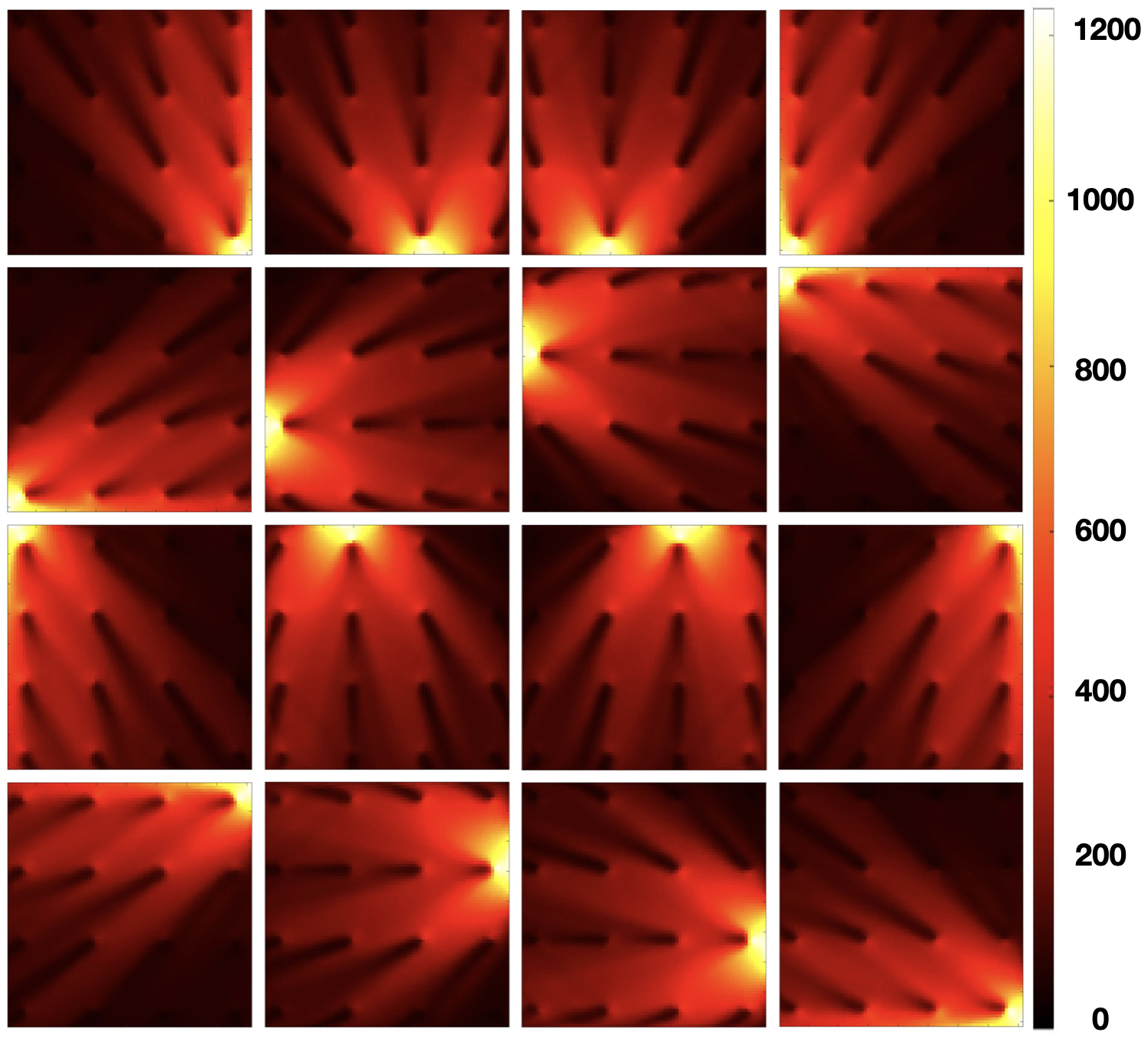}
			\includegraphics[scale=0.23]{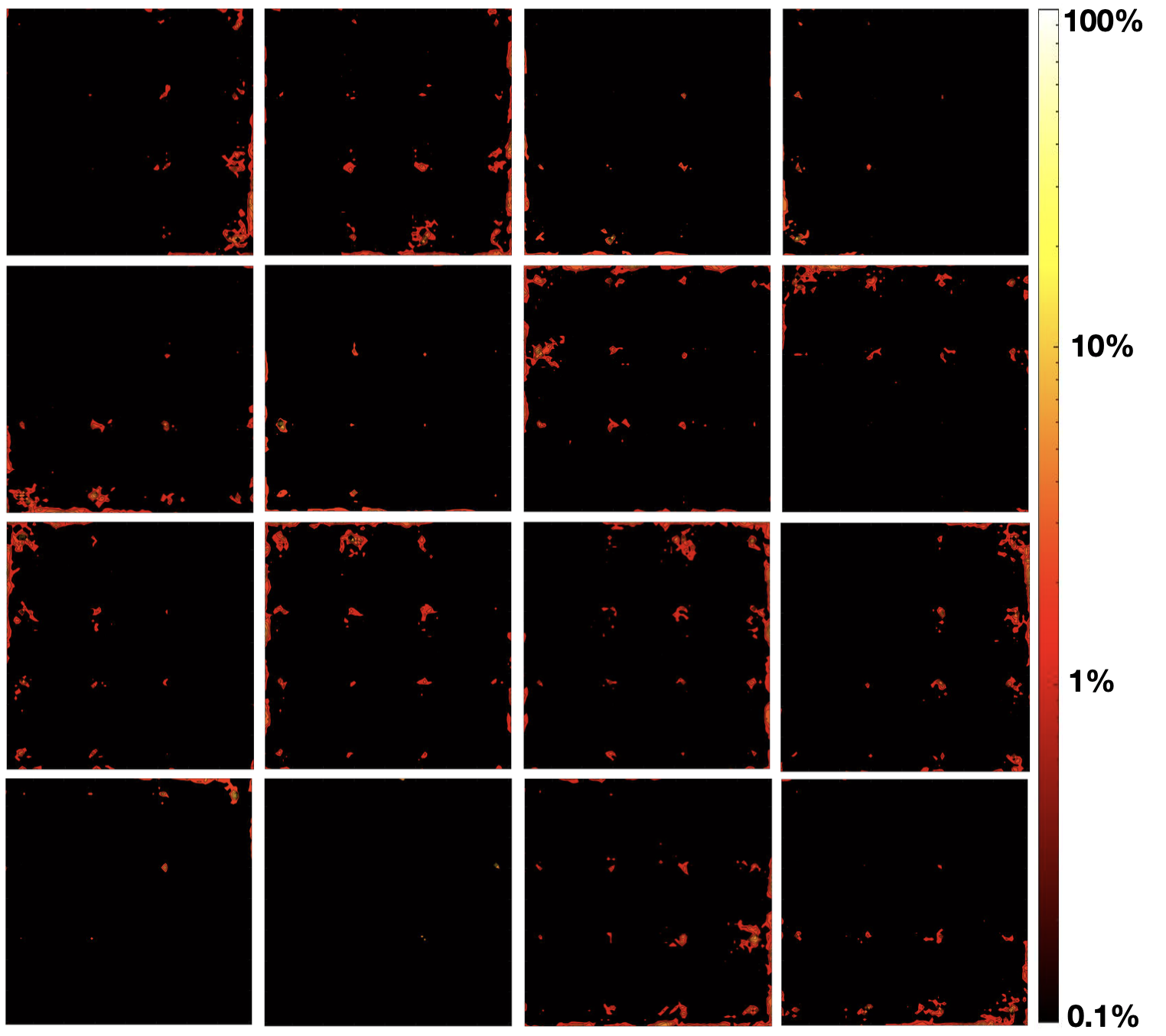}
			\caption{Top: MDRFs achieved using a traditional scanning pencil-beam method. Middle: MDRFs achieved using the CDS method. Bottom: the logarithmic error maps achieved by comparing the top and middle figures (mean and maximum errors are listed in TABLE III, and color bar was set to be the percentage of the maximum MDRF value of the top figure).}
		\end{center}
	\end{figure}\par
	
	\begin{table}[H]
		\fontsize{5}{7}\selectfont
		\centering
		\caption{Mean and maximum value of each error map of Fig. 18 (bottom), as percentage of maximum map value of Fig. 18 (top, ground truth MDRFs).}
		\label{T:peak}
		\includegraphics[scale=0.4]{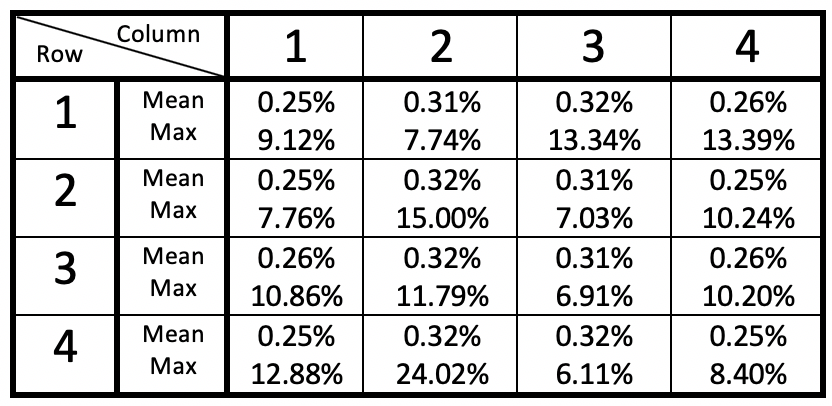}
	\end{table}
	
	After estimating the test events' positions using the MDRFs, the estimated positions were compared with their corresponding true positions. The bias, standard deviation and root mean square error maps are shown in Fig. 19.\par
	
	\begin{figure} [H]
		\begin{center}
			\includegraphics[scale=0.27]{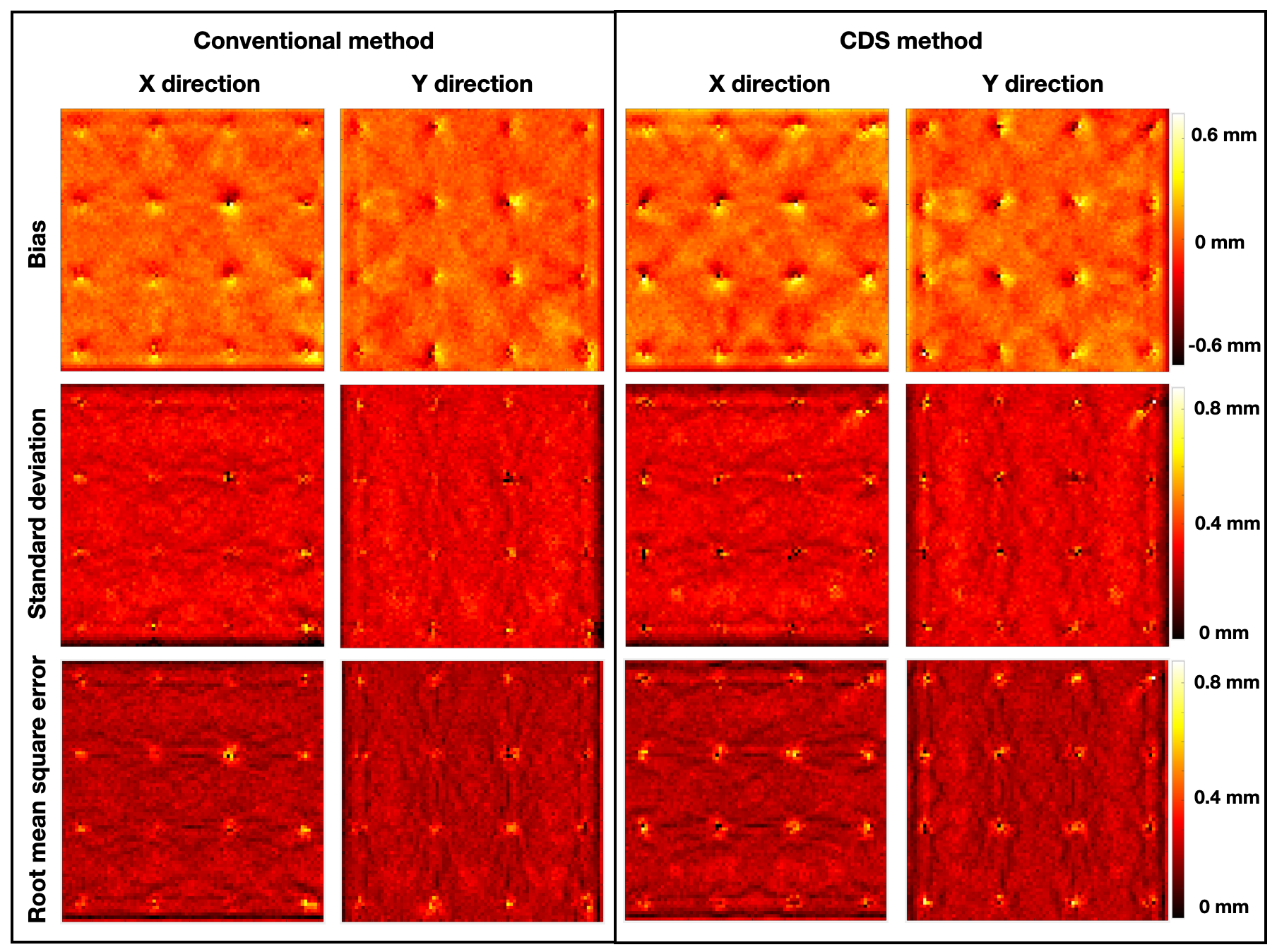}
			\caption{The bias, standard deviation and root mean square error maps of events using MDRFs generated with pencil-beam and CDS methods. For this edge-readout detector, the high error region is where the optical barriers locate, there is no scintillation material. Events assigned to those locations have a lower likelihood value, and could be filtered out using a likelihood threshold.}
		\end{center}
	\end{figure}\par
	
	\subsection{Experimental result for edge-readout detector} 
	The MDRFs acquired by scanning the pencil beam generated with the crossed-slits method are shown in Fig. 20 (top). The MDRFs generated using the CDS method are shown in Fig. 20 (middle). The error maps are shown in Fig. 20 (bottom), with its corresponding mean and maximum value shown in TABLE IV.
	
	\begin{figure} [H]
		\begin{center}
			\includegraphics[scale=0.23]{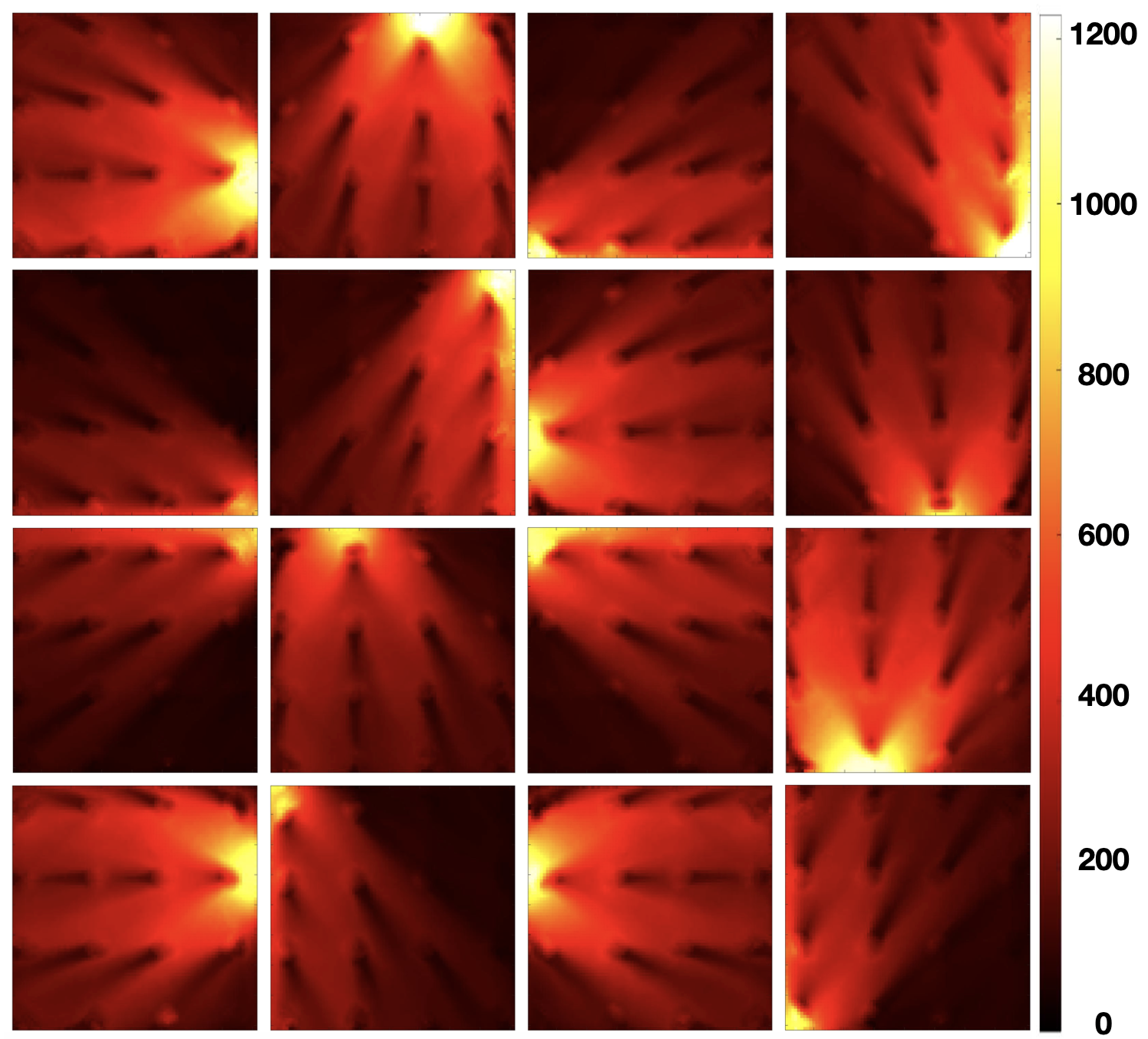}
			\includegraphics[scale=0.23]{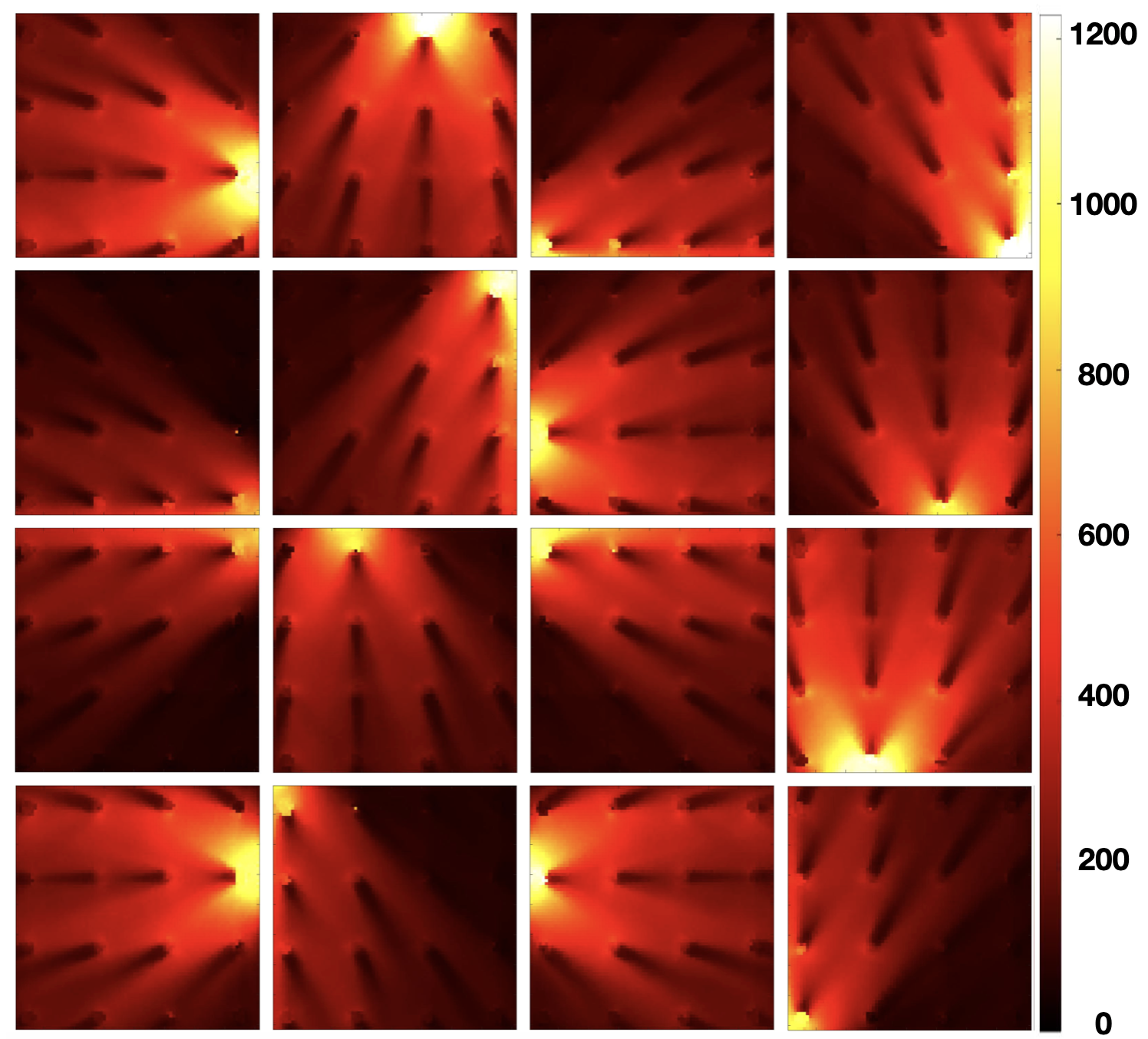}
			\includegraphics[scale=0.23]{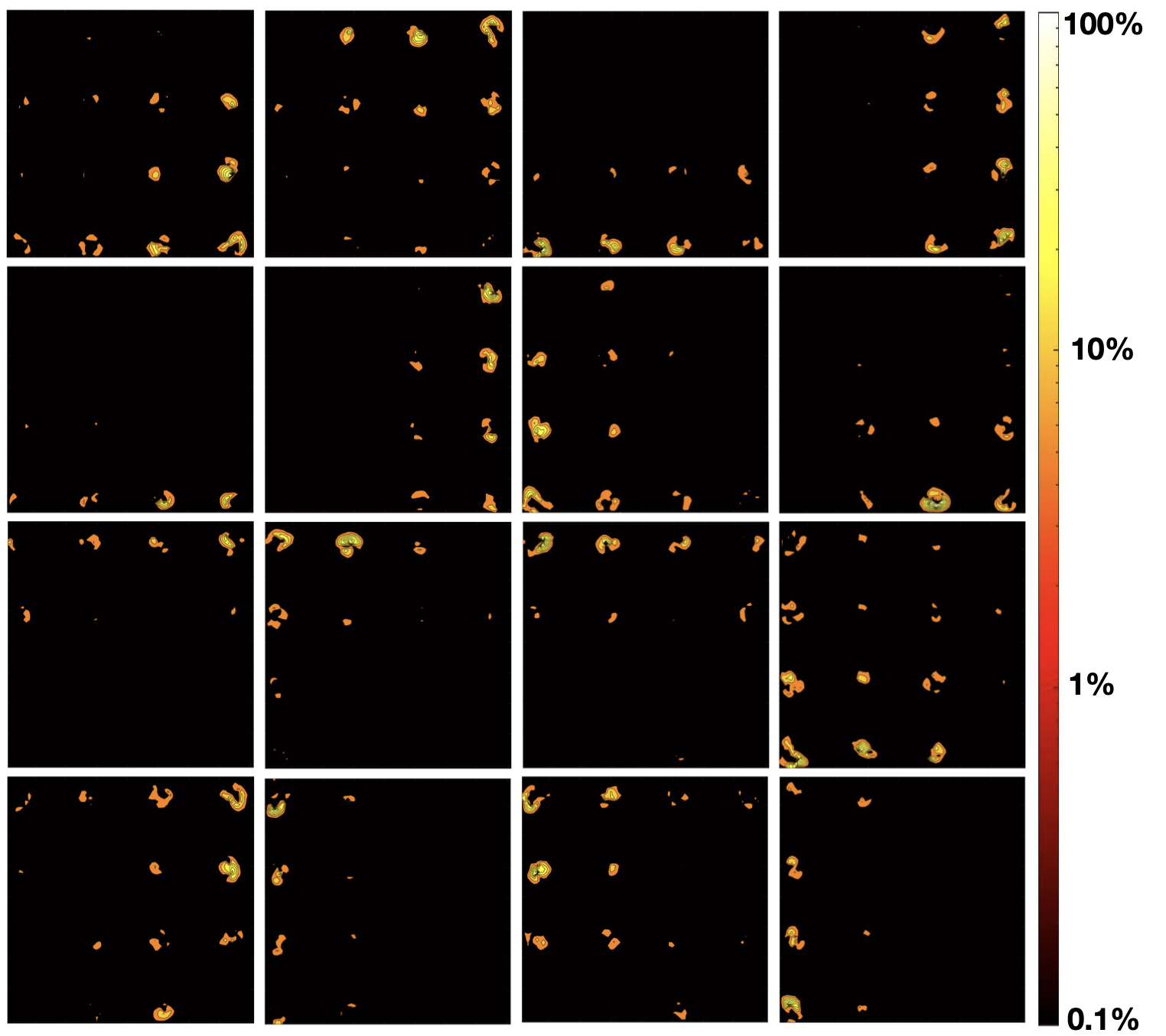}
			\caption{Top: MDRFs achieved from the scanning pencil-beam method. Middle: MDRFs determined using the CDS method. Bottom: the logarithmic error maps achieved by comparing the top and middle figures (mean and maximum errors are listed in TABLE IV) and the color bar was set to be the percentage of the maximum MDRF value of the top figure.}
		\end{center}
	\end{figure}\par
	
	\begin{table}[H]
		\fontsize{5}{7}\selectfont
		\centering
		\caption{Mean and maximum value of each error map of Fig. 20 (bottom), as percentage of maximum map value of Fig. 20 (top).}
		\label{T:peak}
		\includegraphics[scale=0.4]{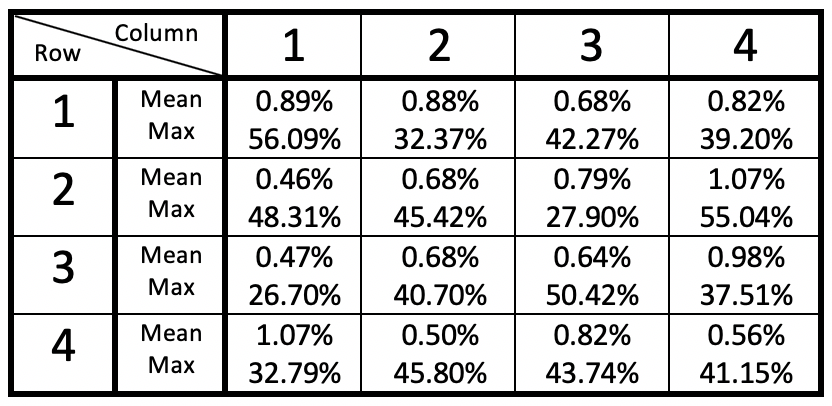}	
	\end{table}
	
	Since it is impossible to use an infinitely thin pencil beam to generate test events experimentally, accurate error maps cannot be generated. Instead, $10$ slit beams were used to quantify the accuracy ($5$ horizontal beams and $5$ vertical beams). The estimated positions of the beam interaction events were binned into images. The images using events estimated with MDRFs achieved by conventional pencil-beam method are shown in Fig. 21, while the ones using events estimated with MDRFs achieved by CDS method are shown in Fig. 22. Note that before using those events for CDS data processing, an energy window was first used to filter out events that deposit partial energy, and the remaining events' positions were estimated with exhaustive search method using likelihood as a metric (Maximum-likelihood estimation method\cite{MLPR}). At last, a likelihood threshold is used to filter out events with a lower likelihood (such as events that scatter multiple times in the detector).\par
	
	If using the each event's estimated position in Fig. 21 as ground truth, the standard deviation of events in Fig. 22 away from events in Fig. 21 along the x and y directions are $128.8$ um and $129.3$ um respectively (by comparing pair-wisely).\par
	
	\begin{figure} [H]
		\begin{center}
			\includegraphics[scale=0.18]{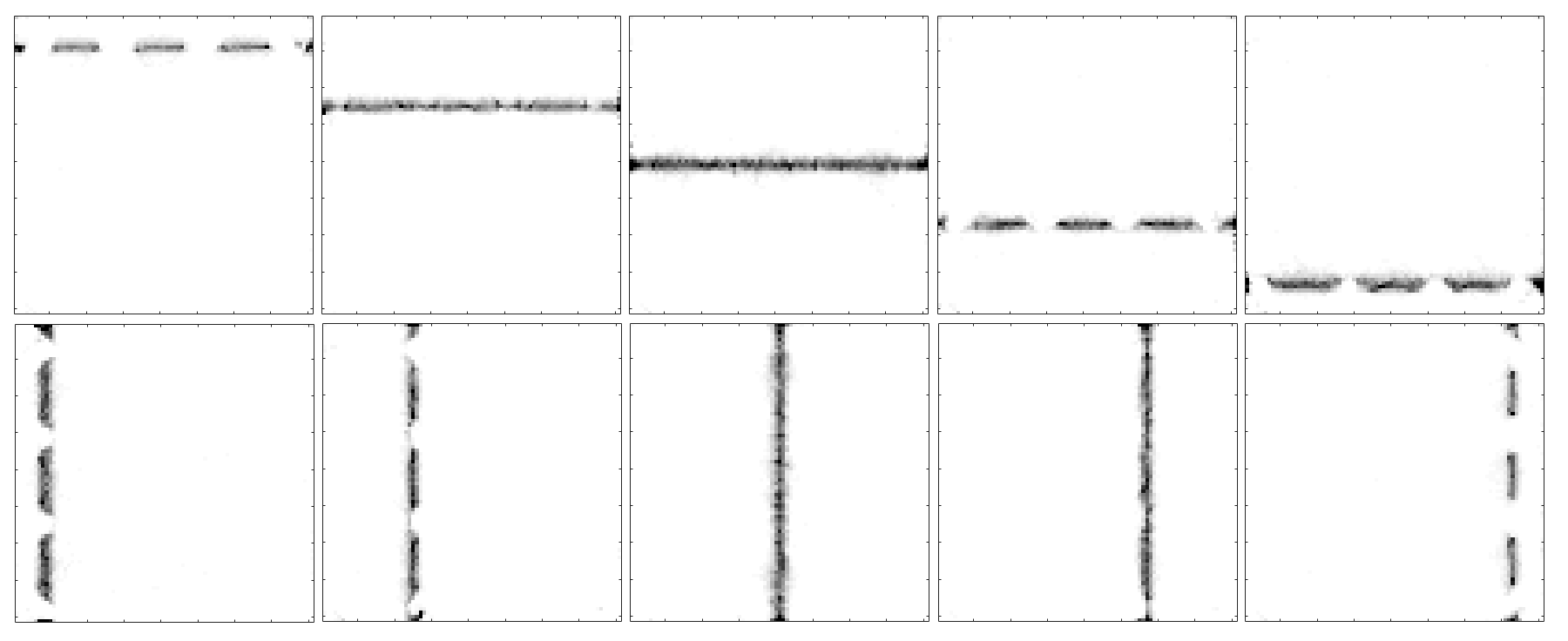}
			\caption{$10$ slit images estimated using the pencil-beam generated MDRFs, the spacing between the slits is $4.0$mm. Top row: $5$ horizontal slit images. Bottom row: $5$ vertical slit images. The disconnections of the slit images are due to the drilled-hole optical barriers, where there is no scintillation material to stop the gamma-ray photons.}
		\end{center}
	\end{figure}\par
	
	\begin{figure} [H]
		\begin{center}
			\includegraphics[scale=0.18]{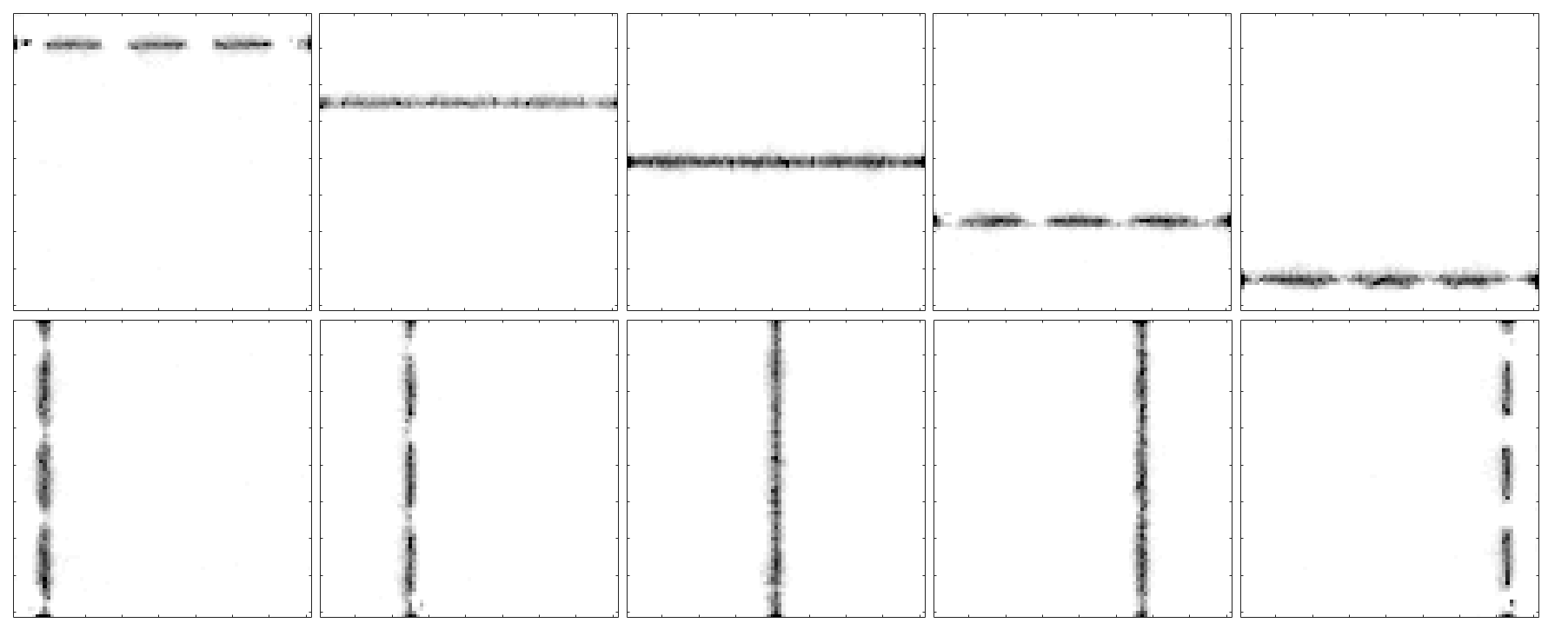}
			\caption{$10$ slit images estimated using the CDS-generated MDRFs, the spacing between the slits is $4.0$mm. Top row: $5$ horizontal slit images. Bottom row: $5$ vertical slit images. The disconnections of the slit images are due to the drilled-hole optical barriers, where there is no scintillation material to stop the gamma-ray photons.}
		\end{center}
	\end{figure}\par
	
	\section{Discussion and conclusion}
	
	The simulation result of the thick monolithic detector shows CDS method's capability of achieving DOI calibration. The maximum MDRF error according to TABLE I (without considering Compton effect) is 3.05\%. Maximum MDRF error according to TABLE II (considering Compton effect) is 20\%, since Compton effect might cause a gamma-ray photon to deposit energy at multiple locations, thus blurring the MDRFs, therefore error is larger at locations with high MDRF gradient. According to the estimated interaction position error maps shown in Fig. 13, the error maps of the ground-truth MDRFs and CDS-generated MDRFs are quite similar. According to Fig. 16, the Compton effect will increase error, which is as expected. However, it is surprising that, comparing Fig. 10 and Fig. 13, the ground-true MDRFs have larger error along z direction at depth of $10$mm compared with CDS-generated MDRFs. In addition, by comparing the error maps of CDS method with and without Compton effect (Fig. 13 and Fig. 17), we can find that the Compton effect will also decrease the DOI error value at depth of $10$mm. As we know that the Compton effect will have blurring/averaging effect on the MDRFs, we suspect the decrease of error effect was caused by the blurring/averaging effect. For CDS method, there will be blurring/averaging effect of MDRFs at regions where the detector resolution is poor. For the depth of $10$mm, the DOI resolution is poor near detector center, so the blurring/averaging effect is induced by the CDS method along Z direction, therefore, the reduction of error phenomenon can be explained, and we have also intentionally introduced blurring to the ground-truth MDRFs by a running average filter, then used these filtered MDRFs to estimate the positions of test events, and created the error maps again. The reduction of DOI error effect is observed again in the same location of the detector. So we concluded that this effect is indeed caused by the blurring/averaging effect. However, this is not to say that adding blurring/averaging effect to MDRFs is always good. The blurring/averaging of MDRFs might only benefit this particular detector design in that specific region along Z direction, while resolution in other part of detector will be negatively affected.\par
	
	The CDS fan-beam calibration method only requires $3$ calibration scans to finish the data collection of a 3D detector, but demands relatively larger amount of computation effort for the CDS data processing. Note that equation (5) is very computationally-consuming, the time complexity is $O(N^2K^2M)$, $N$ is the number of reference datasets of each scanning direction, $K$ represents the number of events in each reference dataset, $M$ represents the number of light sensors or independent readouts in detector. The equations can be calculated in parallel, and GPUs were used to accelerate the computation speed. Moreover, it is found that for equation (5), the score of $\mathbf{a_i}$ can be calculated efficiently by randomly selecting events in dataset $\mathbf{B_{n'}}$ sparsely. For example, the number of events picked from dataset $\mathbf{B_{n'}}$ could be as small as $1/10$ of the total number of events in it, and it further speeds up the computation. We have used a server with 8 Geforce Titan RTX for the CDS data processing, it took about $40$ mins to finish the CDS data process for datasets scanning x direction, with each dataset containing $50,000$ events. After we used the aforementioned sparse sampling method, the computation time was reduced to about $10$ minutes, and we also found that using this method, the result is more sensitive to scatter noise.\par
	
	For the edge readout detector, the simulation results showed good agreement between the the CDS method and pencil-beam method. However, experimentally, we found some discrepancies between the CDS and pencil-beam methods, especially at the regions around drilled-hole optical barriers (as shown in bottom part of Fig. 20), where there is either no scintillation material or the MDRF gradients are relatively large. And due to the error introduced in the iterative filtration process of pencil-beam method aimed to filter out outliers, it is hard to tell which method provides more accurate MDRFs. Despite the difference, the slit images achieved using the MDRFs acquired by CDS method show that the CDS calibration method is capable of generating MDRFs that can accurately estimate gamma-ray interaction positions.\par
	
	It is obvious that the CDS method is not restricted to calibrating scintillation detectors; any detector with photon-counting capability could potentially use CDS methods for fast detector calibration to improve positioning performance.\par
	
	In our future work, we will experimentally implement the CDS calibration method on a large thick monolithic-crystal detector for 3D detector calibration for further validation. We will also explore other algorithms/methods to increase CDS method's data-processing speed.
	
	\appendices
	
	\bibliographystyle{ieeetran}
	\bibliography{referenc}
	
	\ifCLASSOPTIONcaptionsoff
	\newpage
	\fi
	
\end{document}